\documentclass[11pt,letterpaper]{article}
\pdfoutput=1
\usepackage{jheppub}
\usepackage{bbm}
\usepackage{mathrsfs}
\usepackage{slashed}
\usepackage{caption}
\usepackage{epstopdf}
\usepackage[normalem]{ulem}
\usepackage[bottom]{footmisc}
\usepackage{subcaption}
\usepackage{bbold}
\usepackage{titlesec}
\usepackage{threeparttable}
\usepackage{booktabs}
\usepackage{changepage}
\usepackage[utf8]{inputenc}
\usepackage{dsfont} 
\usepackage{float}
\usepackage{placeins}

\usepackage{grffile}

\usepackage{graphicx}  
\usepackage{dcolumn}   
\usepackage{bm}        
\usepackage{amssymb}   
\usepackage{setspace}
\usepackage{amsmath, amssymb, setspace}
\usepackage{array}
\usepackage{booktabs}
\usepackage{caption}
\usepackage{subcaption}
\usepackage{indentfirst} 
\usepackage{float}
\usepackage{lmodern}
\usepackage{multirow}
\usepackage{soul}
\usepackage[normalem]{ulem}
\usepackage{braket}
\usepackage{comment}

\usepackage[draft]{pgf}

\usepackage{adjustbox} 

\newcommand{\ER}{E_{\rm R}}
\newcommand{\Ee}{E_e}
\newcommand{\vmin}{v_{\rm min}}

\newcommand{\ud}{\text{d}}
\newcommand{\sumint}{\int\kern-1.2em\sum}

\newcommand{\Fig}[1]{Fig.~\ref{#1}}
\newcommand{\Eq}[1]{Eq.~(\ref{#1})}

\newcommand{\teta}{\tilde{\eta}}
\newlength{\myimageoversize}
\newsavebox{\myimage}

\usepackage[most]{tcolorbox}

\usepackage{empheq}

\tcbset{colback=white!10!white, colframe=black!50!black, 
        highlight math style= {enhanced, 
            colframe=black,colback=white!10!white,boxsep=0pt}
        }

\titleformat{\subsubsection}
 {\normalfont\fontsize{12}{17}\itshape}{\thesubsubsection}{1em}{}
 
\title{\huge{Halo-Independent Analysis of Direct Dark Matter Detection Through Electron Scattering}}

\author[a]{Muping Chen,}
\author[a]{Graciela B. Gelmini,}
\author[b,a]{and Volodymyr Takhistov}

\affiliation[a]{Department of Physics and Astronomy, UCLA,\\
475 Portola Plaza, Los Angeles, CA 90095, USA}
\affiliation[b]{Kavli Institute for the Physics and Mathematics of the Universe (WPI), UTIAS \\The University of Tokyo, Kashiwa, Chiba 277-8583, Japan}

\emailAdd{mpchen@physics.ucla.edu}
\emailAdd{gelmini@physics.ucla.edu}
\emailAdd{volodymyr.takhistov@ipmu.jp}

\abstract{Sub-GeV mass dark matter particles whose collisions with nuclei would not deposit sufficient energy to be detected, could instead be revealed through their interaction with electrons. 
Analyses of data from direct detection experiments usually require assuming a local dark matter halo velocity distribution. In the halo-independent analysis method, properties of this distribution are instead inferred from direct dark matter detection data, which allows then to compare different data without making any assumption on the uncertain local dark halo characteristics. This method has so far been developed for and applied to dark matter scattering off nuclei. Here we demonstrate how this analysis can be applied to scattering off electrons.}

\begin{document}
\preprint{IPMU21-0030}
 \maketitle
\flushbottom

\section{Introduction}
\label{sec:introduction}

The predominant form of matter in the Universe, the dark matter (DM), has so far only been detected though its gravitational interactions. Its nature remains elusive and multitude of directions have been considered to search for possible non-gravitational DM interactions (see e.g. \cite{Bertone:2004pz} for review). A well studied DM paradigm is that of  Weakly Interacting Massive Particles (WIMPs) with typical mass in the GeV to 100 TeV range which often appear in models that can address the hierarchy problem. But many other DM particle candidates are possible, with mass spanning decades of orders of magnitude. One such possibility is that of DM particles with mass in the sub-GeV range, appearing in a variety of models (e.g.~\cite{Feng:2008ya,Boehm:2003hm,Lin:2011gj,Hooper:2008im,Hochberg:2014dra,Hochberg:2014kqa}).

Direct DM detection attempts to measure the energy deposited within a detector by collisions of DM particles from the dark halo of our Galaxy passing through the detector. The energy deposited on nuclei by DM particles with masses heavier than a GeV can be large enough to be above the detection threshold in most experiments (e.g.~\cite{Cushman:2013zza,Gelmini:2018ogy}). Lighter DM particles could instead be efficiently detected though their scattering off electrons in noble gases~\cite{Essig:2011nj,Graham:2012su,Lee:2015qva,Essig:2017kqs,Catena:2019gfa,Agnes:2018oej,Aprile:2019xxb,Aprile:2020tmw},
semiconductors~\cite{Essig:2011nj,Graham:2012su,Essig:2012yx,Lee:2015qva,Essig:2015cda,Derenzo:2016fse,Hochberg:2016sqx,Bloch:2016sjj,Kurinsky:2019pgb,Trickle:2019nya, Griffin:2019mvc,Griffin:2020lgd,Du:2020ldo}, and superconductors and Dirac materials~\cite{Hochberg:2015pha,Hochberg:2015fth,Hochberg:2016ajh,Hochberg:2017wce,Coskuner:2019odd,Geilhufe:2019ndy}\footnote{See also e.g.~\cite{Gelmini:2020kcu,Lawson:2019brd,Gelmini:2020xir} for other searches.}. Experimental searches of DM scattering off electrons
are currently underway, including dielectric crystal targets, such as Ge (EDELWEISS~\cite{Armengaud:2018cuy, Armengaud:2019kfj, Arnaud:2020svb}, SuperCDMS) and Si (DAMIC~\cite{deMelloNeto:2015mca,Aguilar-Arevalo:2019wdi,Settimo:2020cbq}, SENSEI~\cite{Tiffenberg:2017aac,Crisler:2018gci,Abramoff:2019dfb,Barak:2020fql}, SuperCDMS~\cite{Agnese:2014aze, Agnese:2015nto, Agnese:2016cpb, Agnese:2017jvy, Agnese:2018col, Agnese:2018gze, Amaral:2020ryn}) and noble gas targets, Xe (XENON~\cite{Aprile:2019xxb,Aprile:2020tmw}, LZ~\cite{Mount:2017qzi}) and Ar (DarkSide~\cite{Agnes:2018oej}).

There are two complementary methods to analyse direct DM detection data, the halo-dependent and the halo-independent. The halo-dependent method, employed since the inception of direct detection searches in the 1980's~\cite{Ahlen:1987mn}, requires assuming a model of the local DM  velocity distribution and density. With this input, regions of interest and limits can be obtained in a DM mass-reference cross section $(m, \sigma_{\text{ref}})$ space for a particular type of DM interaction (where the reference cross-section $\sigma_{\text{ref}}$ is a parameter extracted from the scattering cross-section to indicate its magnitude).  

The halo-independent data analysis method does not require assuming a model for the local dark halo. This  avoids the uncertainties associated with our knowledge of the local characteristics of the Galactic halo at the small scales relevant for direct detection, which are much smaller than the scales reached with astrophysical methods. In this method the local DM distribution is inferred from putative DM signals, under the assumption of a particular DM particle model, i.e. given the DM interaction cross section and mass. Distinct data sets can then be compared by their inferred local dark halo properties. 

The halo-independent method has so far been applied to DM collisions off nuclei (see e.g.~\cite{Fox:2010bz,Fox:2010bu,Frandsen:2011gi,Gondolo:2012rs,HerreroGarcia:2012fu,Frandsen:2013cna,DelNobile:2013cta,Bozorgnia:2013hsa,DelNobile:2013cva,DelNobile:2013gba,DelNobile:2014eta,Feldstein:2014gza,Fox:2014kua,Gelmini:2014psa,Cherry:2014wia,DelNobile:2014sja,Scopel:2014kba,Feldstein:2014ufa,Bozorgnia:2014gsa,Blennow:2015oea,DelNobile:2015lxa,Anderson:2015xaa,Blennow:2015gta,Scopel:2015baa,Ferrer:2015bta,Wild:2016myz,Gelmini:2015voa, Gelmini:2016pei,Witte:2017qsy,Gondolo:2017jro,Ibarra:2017mzt,Gelmini:2017aqe,Catena:2018ywo}). However, halo uncertainties can also significantly impact searches of DM scattering off electrons, as
recently stressed in Refs.~\cite{Maity:2020wic,Radick:2020qip}.

Here we demonstrate how the halo-independent analysis can be applied to DM collisions with electrons.  We study elastic scattering off electrons, but our results can be trivially extended to inelastic DM scattering, i.e. scattering in which the incoming and outgoing DM particles have different mass, $m'-m= \delta \ll m$, in the way specified in Sec.~\ref{ssec:inelastic}.

The paper is organized as follows. In Sec.~\ref{Sec-2} we provide a general overview of the halo-independent method. In Sec.~\ref{Sec-3} we apply it to DM interacting with electrons. In particular, we derive the essential element in this formalism, which is the DM particle model and detector dependent response function for scattering off electrons, and compute this function for xenon atoms and semiconductor crystals. Our results are presented and discussed in  Sec.~\ref{sec:resultdisc}. We conclude in Sec.~\ref{sec:conclusion}.

\section{Halo-independent Analysis}
\label{Sec-2}

A halo-independent analysis relies on the separation of the astrophysical parameters contributing to the DM scattering rate, common to all experiments, from the particle physics and detector dependent quantities contributing to the rate. The predicted event rate is written as a convolution  of a function which exclusively depends on the DM velocity distribution, and a kernel we call ``response function" which includes all the rest. The objective of the method is to find the properties of the first, for which it is essential to have the second. 

\subsection{The response function}

In the halo-independent method, direct detection data  are many times translated into measurements of and bounds on a commonly used function we call $\teta(\vmin,t)$, although any other integral of the DM local velocity distribution $f_\chi(\vec{v},t)$ could be used. We call the DM particle $\chi$. The function $\teta(\vmin,t)$ 
\begin{equation}
\label{eq:tilde-eta-t}
\tilde{\eta}(\vmin,t) 
\equiv \frac{\rho~ \sigma_\text{ref}}{m}
\int_{v> \vmin}\ud^3 v \, \frac{f_\chi(\vec{v},t)}{v}
= \frac{\rho~ \sigma_\text{ref}}{m}
\int_{\vmin}^{\infty} \, \ud v \, \frac{F(v,t)}{v}\, 
\end{equation}
is common to all experiments and depends on the speed $\vmin$ defined below. It contains all the dependence of the predicted event rate on the  local dark halo in any direct DM detection experiment (see e.g.~\cite{Fox:2010bz,Fox:2010bu,Frandsen:2011gi,Gondolo:2012rs,HerreroGarcia:2012fu,Frandsen:2013cna,DelNobile:2013cta,Bozorgnia:2013hsa,DelNobile:2013cva,DelNobile:2013gba,DelNobile:2014eta,Feldstein:2014gza,Fox:2014kua,Gelmini:2014psa,Cherry:2014wia,DelNobile:2014sja,Scopel:2014kba,Feldstein:2014ufa,Bozorgnia:2014gsa,Blennow:2015oea,DelNobile:2015lxa,Anderson:2015xaa,Blennow:2015gta,Scopel:2015baa,Ferrer:2015bta,Wild:2016myz,Gelmini:2015voa, Gelmini:2016pei,Witte:2017qsy,Gondolo:2017jro,Ibarra:2017mzt,Gelmini:2017aqe,Catena:2018ywo}). 
In Eq.~\eqref{eq:tilde-eta-t}, $\rho$ is the local DM density, the local DM speed distribution $F(v,t) \equiv v^2 \int \ud \Omega_v f_\chi(\vec{v},t)$  is normalized to 1, $\int_0^\infty{\rm d}v~F(v,t)=1$, as is also the velocity distribution,  $\vec{v}$ is the velocity of the DM particle with respect to the detector,  
$v=|\vec{v}|$ is the DM speed, $\rho$ is the local DM density, $m$ is the DM particle mass and $\sigma_\text{ref}$ is a constant extracted from the scattering cross section to indicate its magnitude. 

The parameter $\vmin$ is the minimum speed the DM particle must have to impart in a collision either a particular recoil energy $\ER$  to a target nucleus, or both $\ER$ and momentum transfer $\vec{q}$ to a target electron. In collisions with nuclei (at rest in the detector), the recoil energy is directly related to $\vec{q}$, thus  $\vmin$ depends only on $\ER= q^2/ 2m_N$, $q=|\vec{q}|$. In collisions with electrons, due to their unknown initial momentum (and also unknown final momentum unless the electron is free in the final state), the relation between the recoil energy and the momentum transfer is lost, thus $\vmin$ becomes a function of the total energy $E_e$ imparted to the electron  and $q$, $E_e = q~\vmin-(q^2/{2m})$ (see Sec.~\ref{Sec-3}). To maintain a unified notation,  we will call here $\ER$ the detectable energy in all instances. Thus, in the case of scattering off electrons in an atom which is ionized as a result of the collision, $E_e = \ER + E_{\rm B}$, where $E_{\rm B}$ is the initial state binding energy. Instead,  $E_e = \ER$ for scattering within a semiconductor crystal. Note that experiments cannot directly  measure the detectable energy, but rather a proxy for it we call $E'$ (e.g. some amount of ionization or a number of photo-electrons). 

The DM particle velocity and speed distributions in Earth's frame, and thus $\teta(\vmin,t)$, are periodic functions of time $t$ due to Earth's rotation around the Sun. A harmonic expansion is usually made
for $\tilde{\eta}(\vmin,t)$,
\begin{equation} \label{eq:tilde-eta-expansion}
\tilde{\eta}(\vmin,t) \simeq \tilde{\eta}(\vmin) + \tilde{\eta}^1(\vmin) \cos(2 \pi (t-t_0)/\text{year}) + \dots~,
\end{equation}
for the speed distribution 
\begin{equation}   \label{eq:F-expansion}
F(v,t) \simeq F(v) + F^1(v) \cos(2 \pi (t-t_0)/\text{year}) + \dots \, ,
\end{equation}
and also for the rate. The first terms of these expansions correspond to time-averages over a year. \Eq{eq:tilde-eta-t} relates the expansion coefficients in \Eq{eq:tilde-eta-expansion} and \Eq{eq:F-expansion}.

For clarity, we review the formalism for DM scattering off nuclei before moving to electron targets. The differential event rate per unit of detector mass  as a function of nuclear recoil energy $\ER$  for a DM particle $\chi$ of mass $m$ scattering off a target nuclide $N$ of mass $m_N$, in a particular experiment is given by
\begin{equation}\label{diffrate-tot}
\frac{\ud R}{\ud \ER} = \sum_N \frac{\ud R_N}{\ud \ER} \, ,
\end{equation}
and the differential rate for each target nuclide $N$ is~(e.g.~\cite{Gelmini:2015zpa})
\begin{equation}\label{diffrate}
\frac{\ud R_N}{\ud \ER} =  \frac{\rho}{m} \sum_N\frac{C_N}{m_N}\int_{v \geqslant \vmin(\ER)} \, \ud^3 \, v \, f_\chi(\vec{v},t) \, v \, \frac{\ud \sigma_N}{\ud \ER}(\ER, \vec{v}) \, .
\end{equation}
Here  $C_N$ is the mass fraction of the nuclide $N$ in a detector, thus $C_N/m_N= N_N$ is the number a target nuclides $N$ in a unit of detector mass, $\ud \sigma_N / \ud \ER$ is the DM-nuclide differential cross section in the lab frame, and for elastic collisions (see \Eq{eq:vmin-delta} for inelastic collisions) 
\begin{equation} \label{vmin-N-elastic}
\vmin=  \sqrt{\frac{m_{N} E}{2\mu_{\chi N}^2}}~,       
\end{equation}
where $\mu_{\chi N}= m~ m_N/ (m+ m_N)$ is the DM-nucleus reduced mass. When the detector includes multiple nuclides $N$, the differential rate is the sum over all of them, as in Eq.~\eqref{diffrate-tot}.

For DM-nucleus  contact interactions due to momentum transfer and velocity-independent interaction operators, such as Spin-Independent interactions, 
the differential cross section is $d\sigma_N/dE_R= \sigma_N(E_R)~m_N/ (2\mu_{\chi N}^2v^2)$, where $\sigma_N(E_R)$ is the total
DM-nucleus cross section. For these cross sections,  the rate takes the  simple form
\begin{equation}\label{diffrate-SI}
\frac{\ud R}{\ud \ER} =   \sum_N \frac{\sigma_N(E_R)}{2 \mu_{\chi N}^2}  \tilde{\eta}(v_{\rm min})
\end{equation}
used by Fox, Liu, and Weiner~\cite{Fox:2010bz}, when they introduced the halo-independent method applied only to Spin-Independent interactions and using differential recoil spectra with a simplified treatment of experimental energy resolutions and form factors to obtain $\tilde{\eta}(v_{\rm min})$. 
In order to extend the method to fully include experimental energy resolutions and efficiencies, as well as nuclear form factors with arbitrary energy dependence~\cite{Gondolo:2012rs}, any isotopic composition of the target~\cite{DelNobile:2013cta}, and to apply it to any type of DM-nucleus interaction~\cite{DelNobile:2013cva} several issues need to be taken into account.
To start with, notice that only for scattering off a single  target nuclide is the relation between $\vmin$ and the nuclear recoil energy $\ER$ unique. Otherwise, one needs to choose whether to treat one or the other as independent variable. 
 If $\ER$ is considered an independent variable, then as mentioned above $\vmin$ is the minimum speed necessary for the incoming DM particle to impart a nuclear recoil $\ER$ to the target nucleus and, thus it depends on the target nuclide $N$ through its mass $m_N$, $\vmin^N=\vmin(\ER,m_N)$. This was the approach in early halo-independent analysis papers (e.g.~\cite{Fox:2010bz}). Alternatively, as we do here, one can chose $\vmin$ as the independent variable, in which case $\ER^{N}(\vmin)$ is the extremum recoil energy (the maximum for elastic scattering, and either the maximum or the minimum for inelastic scattering- see App.~\ref{app:nuclearscat}) that can be imparted to a target nuclide $N$ by an incoming WIMP traveling with speed $v = \vmin$. In this case the recoil energy depends on the target nuclide. Only for scattering off a single  target nuclide are the two approaches related by a simple change of variables. Taking $\vmin$ as independent variable, as we do here, allows one to account for any isotopic target composition by summing terms dependent on $\ER^{N}(\vmin)$ over target nuclides $N$, for any fixed detected energy $E'$.
 
 In fact, as mentioned above, experiments do not actually measure the recoil energy of a target nucleus, but rather a proxy $E'$ for it (e.g. the number of photoelectrons detected in a photomultiplier tube or some amount of ionization or heat). The predicted measured differential rate as a function of the detected energy $E'$ involves a convolution of the recoil rate as function of $\ER$ with the energy resolution function $G_N(E',\ER)$ of the experiment, the function that gives the probability that a detected energy $E'$ resulted from a true recoil energy $\ER$, and also takes into account the efficiency function $\epsilon(E')$ (this is also Eq.~\eqref{eq:diffrate_ep})
\begin{equation}\label{eq:diffrate_withf}
\frac{\ud R}{\ud E'} = \epsilon(E') \sum_N \int_0^\infty \ud \ER  \, G_N(E',\ER) \, \frac{\ud R_N}{\ud \ER} \, .
\end{equation}
Using Eq.~\eqref{diffrate} in Eq.~\eqref{eq:diffrate_withf} and changing the order of the $\vec{v}$ and $\ER$ integrations,  the differential rate as function of the detected energy $E'$ can be written as~\cite{DelNobile:2013cva,Gelmini:2015voa,Gelmini:2016pei,Gondolo:2017jro, Gelmini:2017aqe}
\begin{equation}\label{diffrate_manip1-nuc}
\frac{\ud R}{\ud E'} = \frac{\sigma_\text{ref}~ \rho}{m} \int \ud^3 v \, \frac{f(\vec{v},t)}{v} \, \frac{\ud \mathcal{H}}{\ud E'} (\vec{v}, E') \, ,
\end{equation}
where we define a DM particle candidate and experiment dependent differential response function ${\ud \mathcal{H}_N}/{ \ud E'}$ for every nuclide,  and the total response function is the sum over all nuclides
\begin{equation}\label{eq:dHcurlTotal'-gen}
\frac{\ud \mathcal{H}}{\ud E'}(\vec{v}, E') \equiv \sum_N  \frac{\ud \mathcal{H}_N}{ \ud E'}(\vec{v}, E') \,, 
\end{equation}
and its general expression for  scattering off nuclei is given in \Eq{eq:dHcurl}~\cite{DelNobile:2013cva,Gelmini:2015voa,Gelmini:2016pei}. 

Restricting ourselves to differential cross sections that only depend on the speed of the incoming DM particle $v = |\vec{v}|$, the response function is also only a function of the speed $v$.
In this case, using the speed distribution $F(v,t)$, the rate in \Eq{diffrate_manip1-nuc} can also be written as
\begin{equation}\label{drate_detatilde-nuc}
 \frac{\ud R}{\ud E'} = \frac{\sigma_{\rm ref}\rho}{m}\int_0^\infty{\rm d}v~\frac{F(v,t)}{v}~\frac{{\rm d} \mathcal{H}}{{\rm d}E'}(v, E') .
 \end{equation} 
Then, using the relations,
\begin{equation} \label{derivative-of-CurlyH}
\frac{\ud \mathcal{R}}{\ud E'}(\vmin, E') \equiv \frac{\partial}{\partial \vmin}\left[ \frac{\ud \mathcal{H}}{\ud E'}(\vmin, E') \right] \, ,
\end{equation}
and
\begin{equation} \label{derivative-tilde-eta}
\frac{\sigma_\text{ref}~\rho}{m} \, \frac{F(v,t)}{v} = - \frac{\partial \tilde{\eta}(v,t)}{\partial v} \, ,
\end{equation}
and taking  into account that for $v \to \infty$ $\tilde{\eta}(\infty,t) = 0$, and for $v=0$
\begin{equation} 
\label{curlyHforv=0}
\frac{\ud \mathcal{H}}{\ud E'} (E', 0) = 0 \, 
\end{equation}
(because no event can be produced by a DM particle with $v=0$)~\cite{DelNobile:2013cva,Gelmini:2015voa,Gelmini:2016pei, Gelmini:2017aqe}, 
after an integration by parts of \Eq{drate_detatilde-nuc} we obtain  
\begin{equation}
\label{diffrate_eta}
\frac{\ud R}{\ud E'} = \int_{0}^{\infty} \ud \vmin~ \tilde{\eta}(\vmin, t) \, \frac{\ud \mathcal{R}}{\ud E'}(\vmin,E') \, .
\end{equation}
Namely, the differential rate as function of the detected energy $E'$ can be written~\cite{Gondolo:2012rs, Gelmini:2015voa, Gelmini:2016pei} as the convolution of the halo function $\tilde{\eta}(\vmin,t)$ and a detector and DM particle model dependent ``response function" ${\ud \mathcal{R}}(\vmin,E')/{\ud E'}$. Based on \Eq{derivative-of-CurlyH} we sometimes call $\ud \mathcal{H}/{\ud E'}$ ``integrated response function" to differentiate it from $\ud \mathcal{R}/{\ud E'}$.

\Eq{drate_detatilde-nuc}   and \Eq{diffrate_eta} have been proven for scattering off nuclei in Refs.~\cite{Gondolo:2012rs, Gelmini:2015voa, Gelmini:2016pei}, and we show in Sec.~\ref{Sec-3} and App.~\ref{app:curlyH} that they also apply to scattering off electrons. 

The response  functions ${\ud \mathcal{R}}/{\ud E'}$  and ${\ud \mathcal{H}}/{\ud E'}$ are functions of the DM speed instead of the DM velocity only if the differential cross section is direction independent and the target is isotropic. This occurs when the incoming DM particle and target are unpolarized and the detector is isotropic. This is most common when considering scattering off nuclei, and also applies to the scattering off atomic electrons in a fluid. The scattering off electrons in a crystal is not isotropic, but we will sum the electron form factors over  all directions, which will allow us to still use \Eq{diffrate_eta}.

In the following we are going to concentrate on deriving the essential element in this formalism, the DM particle model and detector dependent response function  $\ud \mathcal{R}/\ud E'$ in Eq.~\eqref{diffrate_eta}, for DM scattering off electrons. This response is only non-zero for an $E'$ dependent speed range, thus it  acts as a ``window function" in $\vmin$ through which a measured rate  can give information on the local dark 
halo~\cite{DelNobile:2013cva,Gelmini:2015voa,Gelmini:2016pei,Gondolo:2017jro, Gelmini:2017aqe}.  Only in the $\vmin$  range in which  this window function is significantly different from zero the halo function $\tilde{\eta}(\vmin,t)$ can be inferred from direct detection data at a particular event energy $E'$. We use differential rates here, but similar equations hold for rates integrated over energy intervals.

So far the response functions ${\ud \mathcal{R}}/{\ud E'}$ were computed only for DM particles scattering off nuclei, for all types of interactions~\cite{DelNobile:2013cva,Gelmini:2015voa,Gelmini:2016pei, Gelmini:2017aqe}.  Here we will derive the response function for the time-average halo function $\tilde{\eta}(\vmin)$ in \Eq{eq:tilde-eta-expansion} for  the elastic scattering of DM particles off electrons in general, and specifically in  Xe, Si and Ge detectors, showing which $\vmin$ range is accessible for these detectors depending on the DM mass and the energy range in which they operate and show how to adapt these results to inelastic scattering in Sec.~\ref{ssec:inelastic}). 

We are going to derive $\ud \mathcal{R}/\ud E'$
for DM scattering off electrons in two alternative ways: 1-  in Sec.~\ref{Sec-3} directly from the rate expression in \Eq{diffrate_eta}, and 2- in App.~\ref{app:curlyH} by deriving first the response function  ${\ud \mathcal{H}}(\vmin,E')/{\ud E'}$ for the speed distribution in   
\Eq{drate_detatilde-nuc}  (see \Eq{CurlyH-App}) and then taking its derivative with respect to the speed (see \Eq{derivative-of-CurlyH}). This latter derivation puts in  evidence the similarities of the response functions for DM elastic scattering off electrons and inelastic endothermic scattering off nuclei, as shown in App.~\ref{app:nuclearscat}.

\subsection{Inference of the local halo velocity distribution}

How to determine the halo function $\tilde{\eta}(\vmin,t)$ or
other integrals of the DM local velocity distributions and how to use them to analyse DM detection data has evolved with time
since the halo-independent method was proposed in 2010~\cite{Fox:2010bz}, and several different proposals have been made.

Initially, the halo-independent method was proposed 
to compare putative signals and upper limits of different direct detection experiments for DM scattering off nuclei, 
using the recoil energy $\ER$ as independent variable and a simplified treatment of experimental energy resolutions and form factors~\cite{Fox:2010bz,Frandsen:2011gi,Frandsen:2013cna}. As explained above, in this case $\vmin$ is a dependent variable, 
which depends on the nuclide mass. Later,  treating $\vmin$ as independent variable allowed to fully take into account the isotopic target composition by summing over target nuclides, for any fixed detected energy $E'$, as well as energy resolutions functions  and nuclear form factors with any energy dependence~\cite{Gondolo:2012rs,DelNobile:2013cta,DelNobile:2013cva}.

Early on, only weighted averages  were obtained for $\tilde{\eta}(\vmin)$ and for the amplitude $\tilde{\eta}^1(\vmin)$ in \Eq{eq:tilde-eta-expansion}, over $\vmin$ intervals where the response function was sufficiently different from zero (see e.g.~\cite{Fox:2010bz,Frandsen:2011gi,Gondolo:2012rs,DelNobile:2013cva}). Recall that $\tilde{\eta}^1(\vmin)$ is the coefficient of the annually modulated component  of the halo function in the harmonic expansion of \Eq{eq:tilde-eta-expansion}. This procedure yields only a poor understanding of the compatibility of various data sets.

Later,  using Karush-Kuhn-Tucker conditions Refs.~\cite{Fox:2014kua, Gelmini:2015voa} showed only for extended likelihoods, i.e. unbinned data, how to determine the unique best-fit average halo function, $\tilde\eta_{BF}(\vmin)$ and proved that this function is  always a piecewise constant non-increasing function with at most $\mathcal{N}-1$ downward steps, where $\mathcal{N}$ is the total number of data entries.  It was also shown how to construct two-sided pointwise confidence bands in the $\vmin-\teta$ plane at any chosen confidence level~\cite{Gelmini:2015voa, Gelmini:2016pei}.
These proofs had strong limitations, since the same procedure could not be applied to binned data or to measurements of modulation amplitudes.  Besides, they provided no  insight into why  $\tilde\eta_{BF}(\vmin)$ had the peculiar functional form just mentioned.  This was later clarified using concepts of convex geometry~\cite{Gondolo:2017jro,Gelmini:2017aqe}, leading to a procedure that can be applied to any type of direct detection data.

 Properties of the  convex set of  DM velocity distribution functions fulfilling a set of conditions imposed by measured average event rates or modulation amplitudes were used in Ref.~\cite{Gondolo:2017jro} to extremize an additional event rate or amplitude. Theorems of convex geometry show that the extremum of this additional rate or amplitude can be obtained with ``extreme distribution functions", which consist of a linear (actually, a convex) combination of delta functions in velocity (or speed). Ref.~\cite{Gondolo:2017jro}  set an upper limit on  the average rate corresponding to a measured modulation amplitude, showing how the latter gives information on the DM local velocity distribution function in the Galactic rest frame, and then profiling a likelihood over the DM velocity distribution.

 Ref.~\cite{Gelmini:2017aqe} considered instead the convex hall of rates generated by the response functions to obtain a similar form (that of the extreme distribution functions) for the DM velocity (or speed) distribution  with which any likelihood can be maximized. E.g. when considering only time-average rates, any likelihood can be maximized with a speed distribution of the form~\cite{Gelmini:2017aqe} 
\begin{equation}\label{eq:Fdeltas}
F(v)=\sum_{h=1}^{\mathcal{N}-1} F_h ~\delta(v-v_h)~,
\end{equation}
where $F_h$ and $v_h$ are parameters and $\mathcal{N}$ is the total number of data entries. The reason is that any set of rates can be written in terms of DM distributions of this form, and any likelihood can always be maximized for a particular set of predicted rates (however, while the best fit rates are always unique, the best fit DM distribution may not be unique). Notice that
\Eq{eq:Fdeltas}  implies that the time-averaged best-fit halo function $\teta_{BF}(\vmin)$ is piecewise constant  with at most $(\mathcal{N} - 1)$ downward steps, since is an integral over a sum of delta functions in speed (see \Eq{eq:tilde-eta-t}), explaining the result that had been previously found. 

When considering coefficients of the harmonic expansion of the time dependent rate other than its average, i.e. modulation amplitudes, the dependence of the rate on the  DM velocity (instead of the speed) needs to be taken into account. In this case it is necessary to change the reference frame from Earth’s frame to the Galactic frame, so that the time dependence of the rate is shifted from the DM velocity distribution $f^{\rm gal}(\vec u)$ (which is now time-independent) to the periodic  response function, i.e. $\ud \mathcal{H}^{\rm gal} (\vec{u}, E', t)/{\ud E'}$ (which is periodic because the detector rotates around the Sun). This allows to apply the same convex geometry theorems to prove that any likelihood can be maximized with 
\begin{equation} \label{Galactic-velocity}
f^{\rm gal}(\vec u)=\sum_{h=1}^{\mathcal{N}-1} f_h^{\rm gal}~\delta^{(3)}(\vec u-\vec u_h)~, 
\end{equation}
where $f_h^{\rm gal}$ and $\vec u_h$ are parameters. Ref.~\cite{Gelmini:2017aqe} showed how to maximize a likelihood with velocity
or speed distributions as in \Eq{Galactic-velocity} or \Eq{eq:Fdeltas}
to find not only the best fit halo function but also either a confidence or a degeneracy band about it. In fact, Ref.~\cite{Gelmini:2017aqe} proved that for  extended likelihoods the  best-fit $\teta$ function is guaranteed to be unique, while for likelihoods depending only on binned data, such as Poisson or Gaussian, the best-fit function may or may not be unique (and showed how to determine if it is or not unique).
Additionally,  it showed how to find either a pointwise confidence band at a particular confidence level about the best fit if it is unique, or a degeneracy band, namely a band containing all degenerate best-fit halo functions, if it is not.

Speed distribution functions consisting of linear combinations of delta functions were also employed in Ref.~\cite{Ibarra:2017mzt}, which used linear programming techniques to make halo-independent comparisons of direct and indirect DM searches. The purpose was to minimize or maximize rates (not likelihoods) or, with some simplifying assumptions, also modulation amplitudes. The formalism of Ref.~\cite{Ibarra:2017mzt} does not attempt to produce halo models compatible with data.

Once the best-fit halo function $\tilde\eta_{BF}(\vmin)$ (with its corresponding uncertainty band) is determined  from the direct detection data of a particular experiment for a given DM particle model,  this halo function  (with its uncertainty) can be used to predict the rate that should be found in any other direct detection experiment if the DM particle model assumed is correct.  If several direct detection experiments have putative DM signals, the compatibility of  distinct data sets can be assessed for each given DM model using their inferred local dark halo properties  (e.g. by comparing them using a global likelihood  as proposed in Ref.~\cite{Gelmini:2016pei}). If the inferred halo properties of different data sets are compatible for a particular DM model and not for others, it would point to the model producing compatibility to be the right one.

In this paper we are not going to carry out any data analysis.
We are going to show how the halo-independent method can be extended to DM scattering off electrons by concentrating on the essential task of computing the response functions.  As mentioned above, 
these act as window functions through which measured rates can give information on the local properties of the DM halo. We are going to concentrate on time average rates, and the response functions for the time-average halo function $\tilde{\eta}(\vmin)$ in \Eq{eq:tilde-eta-expansion}. We will thus assess here the possibility of experiments based on Ge and Si or Xe to explore either different or coincident $\vmin$ ranges for different DM mass values.

\section{Response functions for DM scattering off electrons}
\label{Sec-3}

As paradigms of target materials in which a bound electron either becomes free due to the collision or is excited from a bound state to another, we consider atoms and semiconductor crystals~\cite{Kopp:2009et,Essig:2011nj,Essig:2012yx,Essig:2015cda,Essig:2017kqs,Emken:2019tni}. The total energy lost  by a DM particle of mass $m$ with initial velocity $\vec{v}$ in a collision with the momentum transfer $\vec{q}$ is
\begin{equation}
	\Delta E_{\rm DM} = \frac{m v^2}{2}-\frac{|{m\vec{v}-\vec{q}}|^2}{2 m}~.
\end{equation}

 Let us first consider the ionization of an atom. In this case, the target electron overcoming a binding energy $E_{\rm B}$ jumps from a bound state to a free state with recoil energy $\ER$ observable in a detector. The total energy gained by the electron, $E_e$, is thus, 
\begin{equation} \label{Ee}
	E_e = E_{\rm B} + \ER~.
\end{equation}
Since the electron is bound to an atom, part of this energy goes in principle also into the recoil of the nucleus of mass $m_N$, $E_N = {|\vec{q}|^2}/{2m_N}$.
Thus, $\Delta E_{\rm DM} = E_e + E_N$, from which it results that
\begin{equation}
\Ee=\vec{q}\cdot\vec{v}-\frac{|\vec{q}|^2}{2\mu_{\chi N}} ~,
\end{equation}
where $\mu_{\chi N}$ is the DM-nucleus reduced mass. However,
 the nucleus is much heavier than the DM particles we study, $m_N \gg m$, thus the DM-nucleus reduced mass can be approximated by the DM mass,  $\mu_{\chi N}\simeq m$. This approximation amounts to neglecting $E_N$ and setting
\begin{equation}
\Ee= \Delta E_{\rm DM}=\vec{q}\cdot\vec{v}-\frac{|\vec{q}|^2}{2m}~.
\end{equation}
Calling $\theta_{qv}$ the angle between $\vec{v}$ and $\vec{q}$, the DM particle speed corresponding to the momentum transfer magnitude $q = | \vec{q}|$ and electron energy $E_e=\ER + E_{\rm B}$ is
\begin{equation} \label{v-of-q-Ee}
v =\frac{\ER + E_{\rm B}}{q \cos{\theta_{qv}} }+\frac{q}{2m  \cos{\theta_{qv}}}~,
\end{equation}
whose minimum value is
\begin{equation} \label{vmin-of-q-Ee}
\vmin(q,\Ee)=\frac{\ER + E_{\rm B}}{q}+\frac{q}{2m}~.
\end{equation}

In a semiconductor crystal, the electron is excited from an initial state $i$ to a final state $f$, with a change of energy $\Delta E_{i\rightarrow f}$ and all of this energy is observable. To keep a common notation in the general equations that follow we are going to still  call  $\ER$ the observable energy, even in crystals. Thus, in a crystal 
\begin{equation} 
E_e= \Delta E_{i\rightarrow f}= \ER~,
\end{equation}
which amounts to taking $E_{\rm B}=0$ in Eqs.~\eqref{Ee}, \eqref{v-of-q-Ee} and \eqref{vmin-of-q-Ee}. The kinematics is the same in a crystal as in an atom except for  $E_{\rm B}=0$ because in both cases the energy lost into the crystal or the atom is negligible. 
\begin{figure*}[tb]
\begin{center}
\includegraphics[trim={0mm 0mm 0 0},clip,width=.49\textwidth]{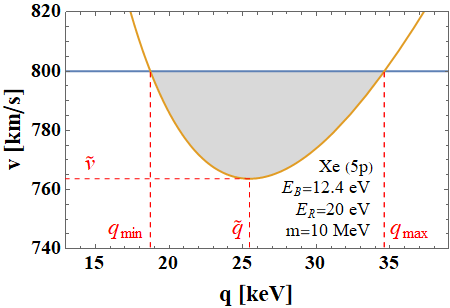} 
\includegraphics[trim={0mm 0mm 0 0},clip,width=.49\textwidth]{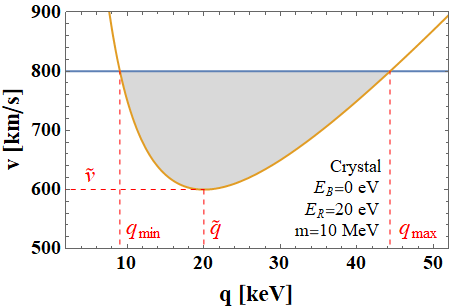} 
\caption{ \label{fig:vqE}
 Function $v_{\rm min}(q, \ER+ E_{\rm B})$ in \Eq{vmin-of-q-Ee} (orange line) and integration range in $q$ in Eq.~\eqref{eq:dRiondER} for $m= 10$ MeV, $\ER= 20$ eV, and  the 5p orbital in a xenon target with $E_{\rm B}=12.4$ eV (left panel)  and  for a semiconductor target thus $\ER= E_e$ (right panel). We call generically $\ER$ the detectable energy in both cases. Shown here are $\tilde{v}$ (the minimum $v_{\rm min}$ value), $\tilde{q} = q_\pm(\tilde{v})$, $q_{\rm min}=q_{-}(v_{\rm max},\ER+ E_{\rm B})$ and $q_{\rm max}=q_{+}(v_{\rm max},\ER+ E_{\rm B})$.  Here $v_{\rm max}=$ 800 km/s, corresponding to DM bound to the Galaxy. } 
\end{center}
\end{figure*}

The function $\vmin(q,E_e)$ in \Eq{vmin-of-q-Ee} is shown in \Fig{fig:vqE} as function of $q$ (orange lines) for $m = 10$~MeV, $\ER= 20$~eV and either $E_{\rm B}= 12.4$~eV (left panel), corresponding to the 5p orbital of a Xe atom,   or $E_{\rm B}= 0$ (right panel), corresponding to electrons jumping between two  bands of a semiconductor crystal, where the whole energy change is observable.  

For fixed $E_e$, $q$ as function of $\vmin$ has two solutions, 
	\begin{eqnarray} \label{qbranches}
	q_{\pm}(\vmin,E_e)=m\vmin\left( 1 \pm \sqrt{1-\frac{2 E_e}{m\vmin^2}} \right)~,
	\end{eqnarray}
which meet at the minimum $\vmin$ value
\begin{equation} \label{def-vtilde}
\tilde{v}=\sqrt{2 E_e/m}~,
\end{equation}
where $q$ has the value
 \begin{equation} \label{def-qtilde}
 q_{\pm}(\tilde{v},E_e)= \tilde{q}=\sqrt{2 m E_e}~,
 \end{equation}
as shown in Fig.~\ref{fig:vqE}.

The maximum possible value of $\vmin$ is the maximum possible speed $v_{\rm max}$ of a DM particle in Earth's frame. The maximum speed of a DM particle bound to the halo of the Galaxy is the escape speed from the Galaxy at the position of the Solar System ($\simeq 550$ km/s  \cite{Piffl:2013mla,Monari:2018,Deason:2019}) plus the speed of the Sun with respect to the Galaxy ($\simeq 240$ km/s, see \cite{Benito:2019ngh} for discussion of uncertainties), which we take to be $v_{\rm max}=$ 800 km/s.  DM that is not bound to the Galaxy, such as DM from the Local Group and the Virgo Cluster, could in principle also contribute subdominantly to direct 
detection~\cite{Baushev:2012dm,Freese:2001hk,Herrera:2021puj}, in which case $v_{\rm max}$ would be larger. We do not consider this possibility, but our formalism can readily accommodate it, by changing the value of  $v_{\rm max}$.

The maximum and minimum values of $q$ for a given $E_e$  (also shown in \Fig{fig:vqE}) are thus
\begin{equation}
\label{qmin-max}
q_{\rm min}=q_{-}(v_{\rm max},E_e)~, ~~~~~ q_{\rm max}=q_{+}(v_{\rm max},E_e)~.
\end{equation}

Using the general formulas for DM-induced electronic transitions in App.~A of Ref.~\cite{Essig:2015cda} (and references there in) we take the cross section for the transition of a given target electron from an initial state $i$ to a final state $f$ to be
\begin{equation}
\sigma v_{i\to f} = 
\frac{\overline \sigma_{e}}{\mu_{\chi e}^2} \int \frac{d^3 q}{4 \pi} \,\delta \Big(\Delta E_{i\to f} + \frac{q^2}{2 m} - q v \cos \theta_{q v} \Big)  \times |F_{\rm DM}(q)|^2 | f_{i\to f}(\vec q \,)|^2 ~ .
\label{itof-cross-section}
\end{equation}
Here the reference cross section $\sigma_{\rm ref} = \overline{\sigma}_e$ is the non-relativistic DM–electron elastic scattering cross section with the momentum transfer $q$ fixed to the reference value $\alpha m_e$ (the characteristic speed of a bound atomic electron is the fine structure constant $\alpha$),
$\mu_{\chi e}$ is the DM-electron reduced mass,  $|f_{i\rightarrow f}(\vec{q})|$ is the electron form factor, $F_{\rm DM}(q)$ is a DM form factor that we take
to be either $1$ or $(\alpha ~ m_e/q)^2$.  
The time-average rate for this transition, obtained with the time average DM velocity distribution $f_\chi(\vec v)$ is
\begin{equation}
R_{i\to f} = 
\frac{\rho}{m} \int d^3 v \, f_\chi(\vec v) \, \sigma v_{i\to f}~ .
\end{equation}
We show in  App.~\ref{Appendix-A} that summing these transition rates over all initial and final electron states,
while including the factor
\begin{equation}
1= \int \ud \ER~ \delta(\ER - \Delta E_{i\to f} +E_{{\rm B }i})
\end{equation}
in the integration to insure that the detectable energy has a fixed value $\ER$, we obtain the differential event rate  (see \Eq{eq: App-A-general-dRdER}),
\begin{eqnarray} 	\label{eq:general-dRdER}
\frac{\ud R}{\ud\ER}=\frac{1}{2\mu^2_{\chi e}}\frac{1}{\ER}\sideset{}{'}\sum_{i,f} \int_{q_{\rm min}}^{q_{\rm max}} \ud q\, q~\tilde{\eta}(\vmin(q,\ER+E_{{\rm B}i}))~ |F_{\rm DM}(q)|^2|f^{i,f}(q,\ER)|^2.
\end{eqnarray}
Here $\tilde{\eta}(\vmin(q,\ER+E_{{\rm B}i}))$ is the time-average of the function defined in \Eq{eq:tilde-eta-t} (see \Eq{eq:tilde-eta-expansion}) with $\sigma_{\rm ref}= \overline{\sigma}_e$, the symbol  $\sideset{}{'}\sum_{i,f}$ indicates the sum over distinct initial and final energies, and we have defined the electron form factor
\begin{eqnarray}   \label{fij-def-Sec3}
|f^{i,f}(q,\ER)|^2= \sum_{\substack{ \rm degen.\\ \rm states}} \ER~\delta(\ER-\Delta E_{i\rightarrow f}+E_{{\rm B}i})|f_{i\rightarrow f}(\vec{q})|^2.
\end{eqnarray}
After performing the summations over  all degenerate states, which include summing over all directions, the electron form factor $|f^{i,f}(q,\ER)|$ is independent of the direction of $\vec{q}$ (as pointed out in Ref.~\cite{Essig:2015cda}). 

Here $E_{{\rm B}i}$ is non-zero only when the electron is bound in the initial state and free in the final state (it is the binding energy of the initial state) and  should be taken to be zero $E_{{\rm B}i}=0$ otherwise (recall that in a crystal $\Delta E_{i\to f}$ is the detectable energy, which to use a common notation we still call $\ER $), 

To bring the rate in \Eq{eq:general-dRdER} into the form in Eq.~\eqref{diffrate_eta} and thus identify the response function, we chose $\vmin$ as  the independent variable instead of $q$ and change the integration variable using $\ud q_{\pm} = J_{\pm}(\vmin, \ER +E_{{\rm B}i})~ \ud \vmin$, with the Jacobian factors
	 \begin{eqnarray} \label{Jacobian}
	 J_{\pm}(\vmin,\ER +E_{{\rm B}i})&=&\frac{\partial q_{\pm}(\vmin, E_{\rm B}+ \ER)}{\partial \vmin} \nonumber \\
	 &=& m \left(1 \pm 
	 \dfrac{1}{\sqrt{1-\dfrac{(\ER +E_{{\rm B}i})}{m~\vmin^2/2}}}\right)
= m \left( 1 \pm \dfrac{1}{\sqrt{1-\dfrac{\tilde{v}^2}{\vmin^2}}} \right).
\end{eqnarray}
Calling $I(q, \vmin, \ER)$ the integrand, the  integral in $q$ in \Eq{eq:general-dRdER} becomes
\begin{eqnarray}
	& &\int_{q_{ \rm min}}^{q_{\rm max}}dq \,I(q, \vmin(q, \ER +E_{{\rm B}i}), \ER) \nonumber \\
	&= &\int_{q_{ \rm min}}^{q_{0}}dq \,I(q, \vmin(q, \ER +E_{{\rm B}i}), \ER)+\int_{q_0}^{q_{\rm max}}dq \,I(q, \vmin(q, \ER +E_{{\rm B}i}), \ER) \nonumber\\ 
	&= &\int^{v_{\rm max}}_{\tilde{v}} d\vmin\, (J_+ - J_-)~ I(q(\vmin, \ER +E_{{\rm B}i}), \vmin,\ER) \, .
\end{eqnarray}
Here $v_{\rm max}$ is the maximum value of $\vmin$ and the negative sign in front of $J_-$  is due to the change of integration order because $q_{\rm min} = q(v_{\rm max})$.
	
Notice that for any fixed value of $\ER$ the two Jacobian factors $J_+$ and $J_-$ in Eq.~\eqref{Jacobian}  diverge at $\vmin=\tilde{v}=\sqrt{2 E_e/m}$.  This would then mean that the window function in $\vmin$ for each $E_{{\rm B}i}+ \ER$ energy diverges at $\tilde{v}$, so that the rate measured at a particular $\ER$ would give information on the $\teta$ function only at the particular $\tilde{v}$ value\footnote{The singularity in the Jacobian here is similar to that appearing in endothermic inelastic scattering off nuclei, for which the kinematics in $\ER$ is formally similar to that in $q$ here, as explained in  App.~\ref{app:nuclearscat}.}. 
However, as we explain below, we will integrate over $E_R$, and as function of $E_R$ the integral of the Jacobian factors is finite, because it is of the form  $\int dx/ x^{1/2}$. Thus the result of any integration in $\ER$ of the Jacobian multiplied by any non-singular function of $\ER$ is finite (since it is always smaller than the integral of the Jacobian multiplied by the maximum value of the function in the integration range).   We do need to integrate over $\ER$ even for unbinned data, because the measured energy $E'$ of an event always corresponds to a range of $\ER$ given by the experimental energy resolution function. In fact, the observable differential spectrum (as mentioned earlier, see \Eq{eq:diffrate_withf})
	\begin{eqnarray} \label{eq:rate-ER-int}
	\frac{\ud R}{\ud E'}= \epsilon(E') \int_0^{E_{\rm max}} \ud \ER\, G(E',\ER) \frac{\ud R}{\ud \ER}~,
	\end{eqnarray}
	 depends on the energy resolution function $G(E',\ER)$ (and also on the counting efficiency $\epsilon(E')$).
\begin{figure*}[tb]   
\begin{center}
\includegraphics[trim={0mm 0mm 0 0},clip,width=.48\textwidth]{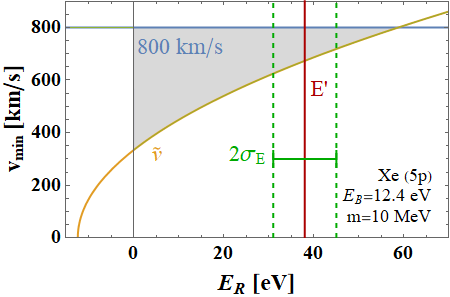} 
\includegraphics[trim={0mm 0mm 0 0},clip,width=.49\textwidth]{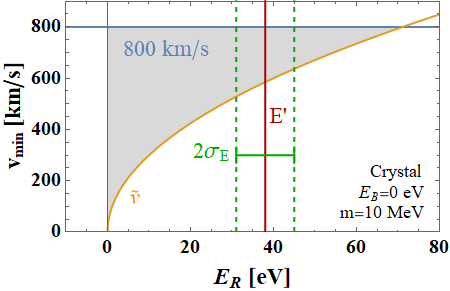} 
\caption{  \label{fig:IntDo}
 Integration domain (in gray) in  $(\ER, \vmin)$ space in \Eq{eq:rate-ER-int} and \Eq{eq:response-ER-int} for $E'= 38$ eV and a box energy resolution function $G(E',\ER)$ with width  $2\sigma_{\rm E}= 14$ eV centered at $E'$, for the 5p orbital of a Xe atom, $E_{\rm B}= 12.4$~eV (left panel) and for a semiconductor target, $E_{\rm B}=0$ (right panel).  Green dashed vertical lines indicate the range in which the energy resolution function  $G(E',\ER)$ is non-zero. The integration range in $\ER$ goes between $E'- \sigma_{\rm E}$ and $E'+ \sigma_{\rm E}$ or between $E'- \sigma_{\rm E}$ and   the orange line $\ER(\tilde{v})$ when  $\ER(\tilde{v})< E'+ \sigma_{\rm E}$. }
\end{center}
\end{figure*}
We can now write the time-averaged differential rate into the form in \Eq{diffrate_eta}, 
	\begin{equation} \label{diffrate_eta-average}
	\frac{\ud R}{\ud E'}=\int^{v_{\rm max}}_{\tilde{v}}  \ud{\vmin}~ \frac{\ud \mathcal{R}}{\ud E'}(\vmin,E')~\tilde{\eta}(\vmin)~,
	\end{equation}
namely as a convolution of the time-averaged $\tilde{\eta}(\vmin)$ in \Eq{eq:tilde-eta-expansion} and the response function,  which we identify to be   
\begin{equation}  \label{kernel-general}
\frac{\ud \mathcal{R}}{\ud E'}(\vmin,E')=   \left(\frac{\ud \mathcal{R}^{+}}{\ud E'}- \frac{\ud \mathcal{R}^-}{\ud E'}\right)~,
	\end{equation}
where     
\begin{eqnarray} \label{eq:response-ER-int}
\frac{d\mathcal{R}_{\pm}}{dE'}(\vmin,E')&=&\frac{\epsilon(E')}{2\mu^2_{\chi e}}\sideset{}{'}\sum_{i,f}\int_{0}^{E_{\rm max}}  \frac{d\ER}{\ER}G(E',\ER)   J_{\pm} (\vmin,\ER+E_{{\rm B}i})~q_{\pm}(\vmin,\ER+E_{{\rm B}i})\nonumber \\
&&|F_{\rm DM}(q_{\pm}(\vmin,\ER+E_{{\rm B}i}))|^2|f^{i,f}(q_{\pm}(\vmin,\ER+
E_{{\rm B}i}),\ER)|^2~.
\end{eqnarray}
 Here, $E_{\rm max}=\frac{1}{2}m\vmin^2-E_{{\rm B}i}$.  The boundaries of the integration domain in the $(\ER,\vmin)$ plane, are easily determined and shown in Fig.~\ref{fig:IntDo}.
In the $(q, \vmin)$ plane, the integration domain is bounded from below by the function $\vmin(q, \ER +E_{\rm B})$ in \Eq{vmin-of-q-Ee},  and from above  by the maximum $\vmin$ value, $v_{\rm max}$. For any $\ER >0 $, there is only one minimum of $\vmin$,  $\tilde{v}$, corresponding to $q=\tilde{q}$. The DM particle speed has to be greater than  $\tilde{v}$ to be detectable. Therefore, in the $(\ER,\vmin)$ plane, for positive $\ER$, the integration domain is bounded by the maximum and  minimum $\vmin$ values, as shown in Fig.~\ref{fig:IntDo}. For each fixed $\vmin$ value in the range $\sqrt{2 E_{{\rm B}i}  /m} \leq \vmin \leq v_{\rm max}$, the $\ER$ integration range is $0 \leq \ER \leq E_{\rm max}=\frac{1}{2}m\vmin^2-E_{{\rm B}i}$. 

\begin{figure*}[tb]
\begin{center}
\includegraphics[trim={0mm 0mm 0 0},clip,width=.49\textwidth]{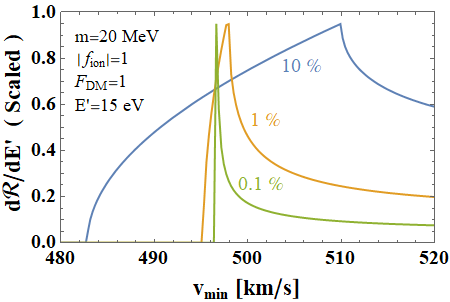} 
\includegraphics[trim={0mm 0mm 0 0},clip,width=.49\textwidth]{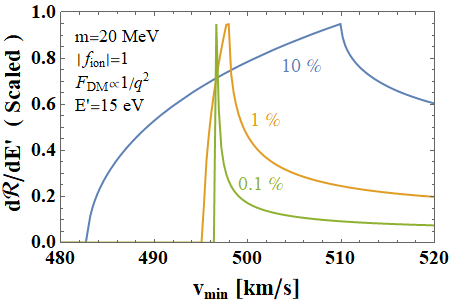} 
\includegraphics[trim={0mm 0mm 0 0},clip,width=.49\textwidth]{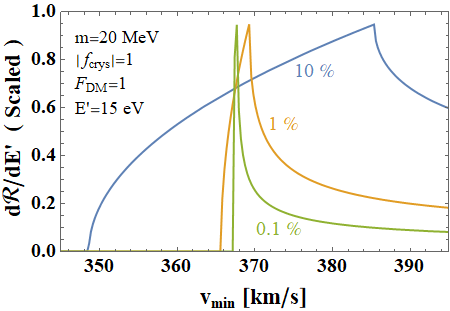} 
\includegraphics[trim={0mm 0mm 0 0},clip,width=.49\textwidth]{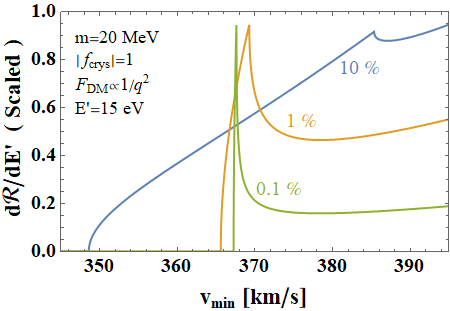}
\caption{ \label{fig:sigmaE} 
Dependence on  the width $\sigma_{\rm E}$ of the energy resolution 
function of the response function $\ud \mathcal{R}/\ud E'$ near  $\vmin =\tilde{v}$ where the Jacobian $J_{\pm}(\vmin,E_{\rm B}+ \ER)$ in Eq.\eqref{Jacobian} has a singularity,  assuming a box resolution function as in \Eq{G-box}, with $\sigma_E$ equal to 0.1\%, 1\% and 10\% of $E'$, for $m$=20 MeV, $E'$=15 eV and electron form factor  set to 1. It shows that a realistic non-zero experimental $\sigma_{\rm E}$ can effectively smooth out the peak.
}
\end{center}
\end{figure*}

Any experimental energy resolution $G(E',\ER)$ has a finite width $\sigma_{\rm E}$. We show in Fig.~\ref{fig:sigmaE} how the shape of the response function's peak (normalized by its maximum value so as to always place the maximum of the ratio close to 1) changes as $\sigma_{\rm E}$ increases, for a box resolution function $G(E',\ER)$ of width $2~\sigma_{\rm E}$ centered at $E'$, $m=$10 MeV and $E'=$ 15 eV, and for simplicity the electron form factor set to 1,  $|f^{i,f}|=1$. Fig.~\ref{fig:sigmaE} demonstrates that the response function  would have a sharp peak as $\sigma_{\rm E} \to 0$, which progressively disappears as $\sigma_{\rm E}$ increases. 

An alternative derivation  of the response function in Eqs.\eqref{kernel-general} and   \eqref{eq:response-ER-int}  based on computing first the response function  $\ud \mathcal{H} (E', v)/\ud E'$ for the DM speed distribution, so that the rate is given by \Eq{diffrate_manip1-nuc}, and then using the relation in \Eq{derivative-of-CurlyH} to compute $\ud \mathcal{R} (E', v)/\ud E'$ as its derivative with respect to the speed $v$ is presented in App.~\ref{app:curlyH}. The function $\ud \mathcal{H} (E', v)/\ud E'$ is given in \Eq{CurlyH-App}. It is regular for all values of the energy and depends on the speed $v$  only though the limits of the integration in $q$. Taking the partial derivative of this function with respect to $v$, and recalling that we defined in \Eq{Jacobian} ${\partial q_{\pm}(v,E_e)}/{\partial v}=J_{\pm}(v,E_e)$,  we recover \Eq{kernel-general} and \Eq{eq:response-ER-int}.

\subsection{Atomic ionization due to DM scattering}

The DM particle can scatter with a bound electron in a particular orbital of an atom ionizing it by taking the electron to an unbound state $\ER=k^2/2m_e$. In an atom target, the initial bound states are labeled by the principal  quantum number $n$ and  the angular momentum quantum number $l$, $|i\rangle=| nl \rangle$,  and the final states are free spherical wave states. The rate given in Refs.~\cite{Essig:2011nj,Essig:2012yx,Essig:2017kqs,Emken:2019tni} is  
\begin{equation}
\label{eq:dRiondER}
\frac{dR_{\rm ion}}{d \ER}=
	\sum_{nl} 
	\frac{1}{8\mu_{\chi e}^2}\frac{1}{\ER}\int_{q_{\rm min}}^{q_{\rm max}} \ud q\,q~\tilde\eta (\vmin(q, \ER + E_{Bnl} )) ~|F_{\rm DM}(q)|^2~|f^{nl}_{\rm ion}(q,\ER)|^2~,
	\end{equation}
	where the limits of integration $q_{\rm min}$ and $q_{\rm max}$ are given in Eq.~\eqref{qmin-max}.  Notice that this coincides with the general rate in Eq.~\eqref{eq:general-dRdER} by identifying the electron form factors, 
	\begin{equation}
	\label{e-form-fact- to- ion}
	\sideset{}{'}\sum_{i,f} |f^{i,f}(q,\ER)|^2= \dfrac{1}{4} \sum_{nl}|f^{nl}_{\rm ion}(q,\ER)|^2~.
	\end{equation}
This identification is done more precisely  in App.~\ref{app:formion}, using the definition of the form factor $\sum_{nl}|f^{nl}_{\rm ion}(q,\ER)|^2$ given in Eq.(6) of Ref.~\cite{Essig:2011nj} (see \Eq{fij-atom}).

\begin{figure*}[tb]
\begin{center}
\includegraphics[trim={0mm 0mm 0 0},clip,width=.49\textwidth]{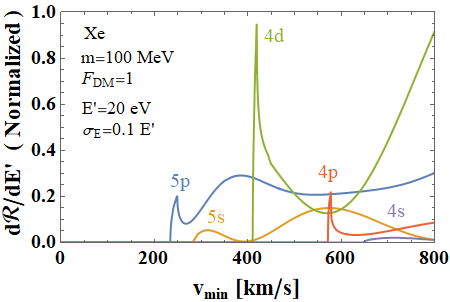}
\caption{ \label{fig:Orb} 
Contribution of several initial state orbitals to the ionization response function ${\ud \mathcal{R}_{\rm ion}}/{\ud E'}$ in xenon (see Eqs.~\eqref{kernel-ion-1}  to \eqref{kernel-ion-nl}), for DM  mass $m=100$ MeV, $E'=20$ eV and  a box energy resolution function with $\sigma_{\rm E}= 0.10 E'$.  In our numerical calculations we only include the dominant 5p and 4d contributions.}
\end{center}
\end{figure*}

Given the experimental energy resolution function of a particular xenon based detector, $G_{\rm ion}(E',\ER)$, the observable DM ionization rate becomes
	\begin{eqnarray}  \label{kernel-ion-1}
	\frac{\ud R_{\rm ion}}{\ud E'}=\int^{v_{max}}_{\tilde{v}}\ud\vmin~	\frac{\ud \mathcal{R}_{\rm ion}}{\ud E'}(\vmin,E')~\tilde{\eta}(\vmin),
	\end{eqnarray}
where the response function is
	\begin{equation}  \label{kernel-ion}
	\frac{\ud \mathcal{R}_{\rm ion}}{\ud E'}= \frac{\ud \mathcal{R}_{\rm ion}^{+}}{\ud E'}-\frac{\ud \mathcal{R}_{\rm ion}^-}{\ud E'}
	\end{equation}
with
	\begin{eqnarray}
	\label{eq:Xekernel}
	\frac{\ud \mathcal{R}^{\pm}_{\rm ion}}{\ud E'}(\vmin,E') 
	&=&\frac{\epsilon(E')}{8\mu^2_{\chi e}}\sum_{nl} \int  \frac{\ud\ER}{\ER}\, G_{\rm ion}(E',\ER)~J_{\pm} (\vmin,\ER+E_{{\rm B}nl})q_{\pm}(\vmin,\ER+E_{{\rm B}nl})
	 \nonumber  \\
&&\times  |F_{\rm DM}(q_{\pm}(\vmin,\ER+E_{{\rm B}nl}))|^2 |f^{nl}_{\rm ion}(q_{\pm}(\vmin,\ER+E_{{\rm B}nl}),\ER)|^2~.
\end{eqnarray}
Here, the electron form factors $|f^{nl}_{\rm ion}|^2$ are those computed in Ref.~\cite{Essig:2015cda} that  we use in our numerical calculations, as explained below in Sec.~\ref{Sec-3.3-numerical}.  
In Fig~\ref{fig:Orb} we show the contribution to the response function of xenon atoms of several initial state orbitals, ${\ud \mathcal{R}^{nl}_{\rm ion}}/{\ud E'}$,  which we defined so that
\begin{equation}  \label{kernel-ion-nl}
	\frac{\ud \mathcal{R}_{\rm ion}}{\ud E'}= \sum_{nl} \frac{\ud \mathcal{R}^{nl}_{\rm ion}}{\ud E'}~.
		\end{equation}
In our numerical calculations (explained in Sec.~\ref{Sec-3.3-numerical}),  we will only use the  4d and 5p  orbitals since they provide the dominant contributions to the response function in xenon.

\subsection{DM scattering in semiconductor crystals}
	
In this case, upon scattering, an electron in the valence band can be excited to a conduction band of the crystal. Since the band gap energy is significantly smaller than the binding energy of orbital electrons, (it is about 1 eV instead of the about 10 eV of ionization threshold in xenon)  crystals have lower threshold and thus greater potential for detecting DM with smaller kinetic energy.  The differential rate given in Refs.~\cite{Essig:2015cda,Emken:2019tni} in this case   is 
	\begin{eqnarray} \label{eq:Essig-rate-crystal}
	\frac{dR_{\rm crys}}{d \ER}=N_{\rm cell}
	\frac{\alpha m_e^2}{\mu_{\chi e}^2} \int dq\,\frac{1}{q^2} \tilde\eta(\vmin(q,\ER))~|F_{\rm DM}(q,\ER)|^2~|f_{\rm crys}(q,\ER)|^2|~,
	\end{eqnarray}
where $N_{\rm cell}$ is number of unit cells in the crystal.

Notice that \Eq{eq:Essig-rate-crystal} coincides with the general rate in Eq.~\eqref{eq:general-dRdER} by taking $E_{{\rm B}i}=0$ and identifying the electron form factor as
	\begin{equation}
	\label{e-form-fact- to- crystal}
\sideset{}{'}\sum_{i,f}	|f^{i,f}(q,\ER)|^2= \dfrac{2 \alpha m_e^2~N_{\rm cell}}{q^3}~|f_{\rm crys}(q,\ER)|^2~.
	\end{equation}
This identification is done more precisely  in App.~\ref{app:formcrystal}, using the definition of the electron form factor $|f_{\rm crys}(q,\ER)|^2$ in a semiconductor crystal in Eq.~(A.33) of Ref.~\cite{Essig:2015cda} (see \Eq{fij-crystal}).  

The rate can now be written as
	\begin{eqnarray}
	\frac{\ud R_{\rm crys}}{\ud E'}= \int^{v_{\rm max}}_{0}\ud\vmin~\frac{\ud \mathcal{R}_{\rm crys}}{\ud E'}(\vmin,E')~\tilde{\eta}(\vmin)~.
	\end{eqnarray}
with the crystal response function
	\begin{equation} \label{kernel-crys}
	\frac{\ud \mathcal{R}_{\rm crys}}{\ud E'}(\vmin,E')=\frac{\ud \mathcal{R}_{\rm crys}^{+}}{\ud E'}-\frac{\ud \mathcal{R}_{\rm crys}^-}{\ud E'}~,
	\end{equation}
where   
	\begin{eqnarray} 
	\label{eq:Gekernel}
	\frac{\ud \mathcal{R}_{\rm crys}^{\pm}}{\ud E'}(\vmin,E')&=&\frac{N_{\rm cell} \epsilon(E') }{\mu^2_{\chi e}} (\alpha m_e^2) \int_0^{E_{\rm max}} \ud\ER  \,G_{\rm crys}(E',\ER) \dfrac{J_{\pm}(\vmin,\ER)}{q_{\pm}^2(\vmin,\ER)} \nonumber \\
	&&\times  |F_{\rm DM}(q_{\pm}(\vmin,\ER))|^2 ~ |f_{\rm crys}(q_{\pm}(\vmin,\ER),\ER)|^2.
	\end{eqnarray}
	Again, in this case the electron form factors are those computed in Ref.~\cite{Essig:2015cda} that we use in our numerical calculations, as explained below. 
 
\subsection{Numerical evaluations of response functions}
\label{Sec-3.3-numerical}

We now describe a general procedure for numerically evaluating the response functions using already provided electron form factor $|f^{i,f}(q,\ER)|^2$ data in energy bins of width $\Delta \ER$ and momentum transfer bins of width $\Delta q$.  
The integrals we need to compute in \Eq{eq:Xekernel} for xenon and  in \Eq{eq:Gekernel} for semiconductor crystals, are discretized into  partitions small enough for the electron form factor in each to be taken as a constant, and the remaining integral is calculated analytically. All the relevant information about the target material electronic structure is contained within the dimensionless form factor that is independent of any physics related to DM. For the crystal form factor $|f_{\rm crys}|^2$ we employ the output data from the \textsf{QEdark}~\cite{Essig:2015cda},  a module for \textsf{Quantum Espresso}~\cite{Giannozzi:2009} based on density functional theory\footnote{For other computational approaches, see e.g.~Ref.~\cite{Griffin:2021znd}.}. For the xenon form factor $|f_{\rm ion}|^2$ we use the data output from Ref. \cite{Essig:2015cda},  in which electronic wave functions are computed assuming a spherical atomic potential and filled electron shells.

To compute the response function in Eq.~\eqref{eq:response-ER-int} (specifically to compute the functions in \Eq{eq:Xekernel} and \Eq{eq:Gekernel}), we need to specify the energy resolution function, the DM form factor, and electron form factors.
 For simplicity, we assume a simple box function for the energy resolution,\footnote{The unit box function $\textrm{UnitBox}[x]$ is 1 for $|x| \leq \dfrac{1}{2}$ and 0 otherwise.}
\begin{equation} \label{G-box}
 G (E', E) = \dfrac{1}{2 \sigma_{\rm E}} \textrm{UnitBox}\Big[\dfrac{\ER - E'}{2 \sigma_{\rm E}}\Big]~.
\end{equation}
With a more realistic Gaussian distribution,  the results are very similar. As can be seen from Eq.~\eqref{eq:response-ER-int}, the resolution width $\sigma_{\rm E}$ affects the limits of integration in energy, thus summation range when the energy range is discretized to perform numerical evaluations. Fig.~\ref{fig:sigmaE} shows how the value of $\sigma_{\rm E}$
affects the form of the response function close to the singularity  point of the Jacobian.  For the other figures we assumed  the resolution $\sigma_{\rm E}=0.10 E'$. 
    
 As already mentioned, following Ref.~\cite{Essig:2015cda} we consider
 two DM form factors, $F_{\rm DM}(q)=1$ or  $F_{\rm DM}(q)=  ({\alpha m_e}/{q})^2$,
which generically appear in a variety of models such as scenarios of vector-portal DM with a dark photon mediator or magnetic-dipole-moment interactions.

\begin{figure*}[tb]
\begin{center} 
\includegraphics[trim={0mm 0mm 0 0},clip,width=.49\textwidth]{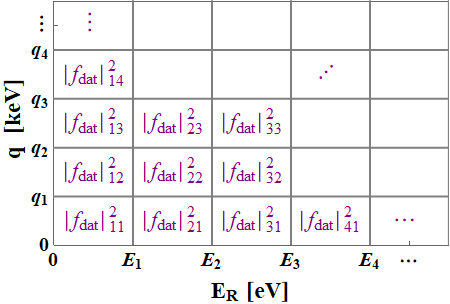}
\includegraphics[trim={0mm 0mm 0 0},clip,width=.49\textwidth]{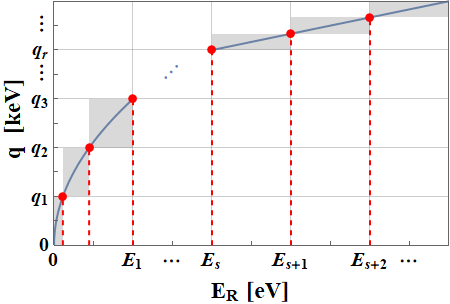}
\caption{ \label{fig:partition}
Discretization mesh of the electron form factor data $|f_{\rm dat}|^2_{rs}$ (left panel) and an example of response function integration path  $q(\vmin,\ER)$ for fixed $\vmin$,  in the $(\ER, q)$ plane.
}
\end{center}
\end{figure*}

We discretize the $(\ER,q)$ plane into a mesh of partitions labelled by $(E_s,q_r)$.  The form factor can then be written as $q_{\pm}^a\ER^b|f^{i,f}_{\rm dat}|^2_{rs}$, where $|f^{i,f}_{\rm dat}|^2_{rs}$ is a given data entry.
For crystals, the data is already provided in bins. For atoms, the table provided was interpolated to find the value at the lowest $q$, $E$ in each partition (and assign it to the partition). Hence, the function in the $(\ER,q)$ plane can be written as a matrix $|f_{\rm dat}|^2_{rs}$, as shown in Fig.~\ref{fig:partition}. 

To perform the integration for the response function numerically, our procedure is the following. Since we are integrating over the path in the $(\ER,q)$ plane described by the function $q(\vmin,\ER)$ at a given $\vmin$, we only need to count the binned partitions that the function passes through. As illustrated on Fig.~\ref{fig:partition}, for some given form factor data (e.g. semiconductor crystals~~\cite{Essig:2015cda}), for each $\Delta E$ bin more than one $\Delta q$ bin might need to be considered. Hence, we use the average value of the electron form factor, weighted by the area of the shaded region in each $q$ interval shown in Fig.~\ref{fig:partition}. This involves finding the $\ER$ value at the intersection of the path taken by $q(\vmin,\ER)$  with each of  the $q$ partition boundaries, which is done via Eq.~\eqref{vmin-of-q-Ee} for a fixed $\vmin$. The resulting $\ER$ points divide $\Delta E$ into smaller sub-intervals, for which we use the corresponding $|f_{\rm dat}|^2_{rs}$ values. Finally, we sum over all possible energy bins within the given $(E'-\sigma_{\rm E},E_{\rm max})$ range.
 
Extracting the  electron form factor data value $|f^{i,f}_{\rm data}(\ER,q_{\pm})|^2$ in each $(E_1^{(rs)},E_2^{(rs)})$ bin from the definition of the response function in Eq.~\eqref{eq:response-ER-int},  we compute analytically the remaining integral, i.e. we write  the response function for each bin as
\begin{eqnarray} \label{eq:general-I-factor}
 \frac{d\mathcal{R}_{\pm}}{dE'}&&(\vmin, E_1^{(rs)},E_2^{(rs)}) = \frac{\epsilon(E')}{2\mu^2_{\chi e}}
 |f^{i,f}_{\rm dat}(E_1^{(rs)},q_{\pm})|^2
\int_{E_1^{(rs)}}^{E_2^{(rs)}}  \frac{d\ER}{\ER}~ q_{\pm}^a\ER^b~G(E',\ER)
\nonumber\\  && \times
 | J_{\pm} (\vmin,\ER+E_{{\rm B}i})   q_{\pm}(\vmin,\ER+E_{{\rm B}i})
|F_{\rm DM}(q_{\pm}(\vmin,\ER+E_{{\rm B}i}))|^2.
\end{eqnarray}
Notice the $q_{\pm}^a\ER^b$ factor, stemming from the fact that the electron form factor in terms of the data values provided in Ref.~\cite{Essig:2015cda} is $|f^{i,f}(q_{\pm},\ER)|^2=q_{\pm}^a\ER^b|f^{i,f}_{\rm data}(\ER,q_{\pm})|^2$, where   $a=b=0$ for atoms and $a=-3$ and $b=1$ for crystals. Then, writing $|F_{\rm DM}(q)|^2 = (\alpha m_e)^{2c}q^{-2c}$, with either $c=0$ or $c=2$, we evaluate the entire response function. The expression for the response function in terms of the $a$, $b$ and $c$ constants is given in the App.~\ref{app:numres} (see \Eq{eq:dRint} of App.~\ref{app:numres}).

For atoms (corresponding to $a=b=0$ in Eq.~\eqref{eq:genres}), the response function is 
\begin{eqnarray} \label{eq:numresion}
\frac{\ud\mathcal{R}^{\pm}_{\rm ion}}{\ud E'}(\vmin, E')=\frac{\epsilon(E')}{16\mu_{\chi e}^2\sigma_{\rm E}}(\alpha m_e)^{2c}\sum_{nl}\sum_{{\rm Path}(r,s)}I^{\pm,nl}_{\rm ion}(\vmin, E_1^{(rs)},E_2^{(rs)})|f_{\rm ion}^{nl}|^2_{rs}~.
\end{eqnarray}
Here, the summation is over all the $(r,s)$ partitions along the path in the $(\ER, q)$ plane of the function $q(\vmin,\ER)$ for fixed $\vmin$, as indicated in \Fig{fig:partition}, and 
\begin{eqnarray}
 I_{\rm ion}^{\pm,nl}(\vmin, E_1^{(rs)},E_2^{(rs)}) \simeq \int_{E_1^{(rs)}}^{E_2^{(rs)}}\ud \ER\, \frac{q_{\pm}^{1-2c}(\vmin, \ER+E_{{\rm B}nl})\,J_{\pm}(\vmin, \ER+E_{{\rm B}nl})}{(E_1^{(rs)}+E_2^{(rs)})/2}.
 \end{eqnarray}
 Thus,
 \begin{eqnarray} \label{Ipm-ion-sec3}
 I_{\rm ion}^{\pm,nl}(\vmin, E_1^{(rs)},E_2^{(rs)}) \simeq
        \begin{cases}
     \dfrac{\mp   (m\vmin)^3}{3(E_1^{(rs)}+E_2^{(rs)})} \left.\left[ \left(  1\pm\sqrt{1-\dfrac{2(\ER+E_{{\rm B}nl})}{m \vmin^2}}\right)^{3}
      \right]\right|_{E_1^{(rs)}}^{E_2^{(rs)}}~,&{\rm if}~c=0;\\
     \\
     \dfrac{\pm  (m \vmin)^{-1}}{(E_1^{(rs)}+E_2^{(rs)})} \left.\left[ \left(  1\pm\sqrt{1-\dfrac{2(\ER+E_{{\rm B}nl})}{m \vmin^2}}\right)^{-1} \right]\right|_{E_1^{(rs)}}^{E_2^{(rs)}}~,&{\rm if}~c=2~.
      \end{cases}
\end{eqnarray}

For semiconductor targets (corresponding to $a=-3$ and $b=1$  in Eq.~\eqref{eq:genres}), the response function is
\begin{eqnarray} \label{eq:numrescrys}
\frac{\ud\mathcal{R}_{\rm crys}^{\pm}}{\ud E'}(\vmin, E') =\frac{\epsilon(E')N_{\rm cell}(\alpha m_e^2)}{2\mu_{\chi e}^2\sigma_{\rm E}}(\alpha m_e)^{2c}\sum_{{\rm Path}(r,s)}I^{\pm}_{\rm crys}(\vmin, E_1^{(rs)},E_2^{(rs)})|f_{\rm crys}|^2_{rs},
\end{eqnarray}
where,  
\begin{eqnarray}  \label{Ipm-crystal-sec3}
 I_{\rm crys}^{\pm}(\vmin, E_1^{(rs)},E_2^{(rs)}) &\simeq& \int_{E_1^{(rs)}}^{E_2^{(rs)}}\ud \ER\, q_{\pm}^{-2-2c}(\vmin, \ER)J_{\pm}(\vmin, \ER)\nonumber \\
 &=&
    \begin{cases}
     \mp \left.\left[\ln\left| 1\pm\sqrt{1-\dfrac{2\ER}{m \vmin^2}} \right|\right]\right|_{E_1^{(rs)}}^{E_2^{(rs)}}~,&{\rm if}~ c=0~;\\
     \\
     \dfrac{\pm 1}{4(m\vmin)^4} \left.\left[ \left(  1\pm\sqrt{1-\dfrac{2\ER}{m \vmin^2}}\right)^{-4} \right]\right|_{E_1^{(rs)}}^{E_2^{(rs)}}~,&{\rm if}~ c=2~.
     \end{cases}
\end{eqnarray}

 In our numerical evaluations, for Si and Ge crystals we use the \textsf{QEdark}~\cite{Essig:2015cda} output data for the computed electron form factors, given in the 0 to 50 eV energy range and 0 to $18~ \alpha m_e$ momentum range with $\Delta \ER=0.1$ eV and $\Delta q=0.02\, \alpha m_e$.  For Xe atoms, we use the same binning and the data table provided in Ref.~\cite{Essig:2015cda} for the electron form factors in the 0.2 to 900 eV energy range and the $0.37~ \alpha m_e= 1.4$ keV to $54.60 ~ \alpha m_e = 203.7$ keV momentum range. The maximum of this range is smaller than $q_{\rm max}= q_{+} (800 {\rm km/s})$ except for the lightest DM masses we consider. 
The table was interpolated to find the form factor value at the lowest $\ER$ and $q$ point in each bin, which was assigned to the bin.

The response functions depend on the DM particle mass $m$, the detected energy $E'$, and the minimum speed $\vmin$. For our figures we chose a set of values for energies  $E'=$ 15 eV, 30 eV and 45 eV and additionally $E'=$ 5 eV for semiconductors (in Xe, the minimum detectable energy is 13.8 eV) 
and DM masses $m=$ 20  MeV, 100  MeV, 1  GeV  that are within the reach of current and near-future experiments\footnote{The average energy needed to produce a single electron quantum is around $\sim$few eV for semiconductors~(e.g.~\cite{Essig:2015cda}) and $\sim$10-15 eV for xenon~(e.g.~\cite{Essig:2017kqs}).}, including
Ge-based EDELWEISS~\cite{Armengaud:2018cuy, Armengaud:2019kfj, Arnaud:2020svb}, SuperCDMS and Si-based DAMIC~\cite{deMelloNeto:2015mca,Aguilar-Arevalo:2019wdi,Settimo:2020cbq}, SENSEI~\cite{Tiffenberg:2017aac,Crisler:2018gci,Abramoff:2019dfb,Barak:2020fql}, SuperCDMS~\cite{Agnese:2014aze, Agnese:2015nto, Agnese:2016cpb, Agnese:2017jvy, Agnese:2018col, Agnese:2018gze, Amaral:2020ryn} and 
Xe-based~\cite{Essig:2011nj,Graham:2012su,Lee:2015qva,Essig:2017kqs,Catena:2019gfa,Agnes:2018oej,Aprile:2019xxb,Aprile:2020tmw} experiments.

 
\subsection{Inelastic DM electron scattering and other possibilities}
\label{ssec:inelastic}

In the case of inelastic electron-DM scattering the DM is multi-component and the initial DM particle of mass $m$ scatters into another of mass $m'$. The mass difference is $\delta = m' - m\ll m\simeq m'$. Going through the same steps we followed in Sec.~\ref{Sec-3}, from the kinematics of the collision we obtain
\begin{equation}
    v_{\rm min} = \dfrac{E_{\rm R} + E_{\rm B}}{q} \Big(1 + \dfrac{\delta}{m}\Big)  +  \dfrac{q}{m}
    = \Big(\dfrac{E_{\rm R} + E_{\rm B}}{q} +  \dfrac{q}{2 m'}\Big) \Big(1 + \dfrac{\delta}{m'}\Big)~. 
\end{equation}
Hence, our halo-independent DM-electron elastic scattering analysis can be readily adapted to inelastic scattering by either scaling   $(E_{\rm R} + E_{\rm B})$ to $(E_{\rm R} + E_{\rm B})(1 + \delta/m_1)$ for the same $v_{\rm min}$ or, alternatively,  scaling  the speed $v_{\rm min}$ to $v_{\rm min}/(1 + \delta/m_2)$ for the same $(E_{\rm R} + E_{\rm B})$.

We have focused on evaluating the response function associated with sub-keV electron signals, but DM-electron scattering can lead to keV signals as well. This has been recently highlighted in relation to the observed XENON1T excess~\cite{Aprile:2020tmw} in the context of inelastic DM-electron scattering (e.g. \cite{Harigaya:2020ckz}) of halo DM particles as well as boosted DM-electron scattering\footnote{Since DM boosted to high velocities is beyond the standard contributions of DM halo, such contribution cannot be used to infer the halo DM distribution using the halo-independent analysis.} (e.g. \cite{Kannike:2020agf}). 
Also keV-level signals could appear from scattering off the extended tail of the momentum distribution of bound electrons (see e.g. lepto-philic DM as discussed in the context of the claimed DAMA experiment signals \cite{Bernabei:2007gr,Kopp:2009et}).

Many of the DM models proposed to explain the XENON1T excess rely on absorption of DM particles~(e.g.~\cite{Aprile:2020tmw}) instead of scattering, in which case the formalism here cannot be applied.  But, if the explanation in terms of scattering off electrons of halo DM particles holds, then as explained in Sec.~2.2, the best-fit halo function $\tilde\eta_{BF}(\vmin)$ (and its corresponding uncertainty band) could be determined using the  XENON1T excess data and taken as the halo model to predict the rate that should be found in any other direct detection experiment, e.g. SENSEI (assuming the DM particle model is correct).

\begin{figure*}[tbh]
\begin{center}
\includegraphics[trim={0mm 0mm 0 0},clip,width=.49\textwidth]{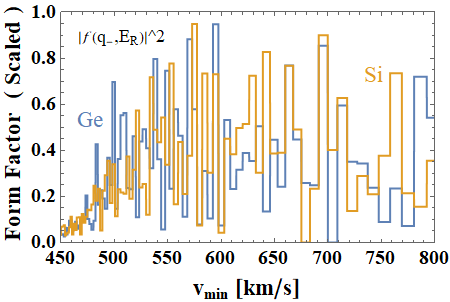}
\includegraphics[trim={0mm 0mm 0 0},clip,width=.49\textwidth]{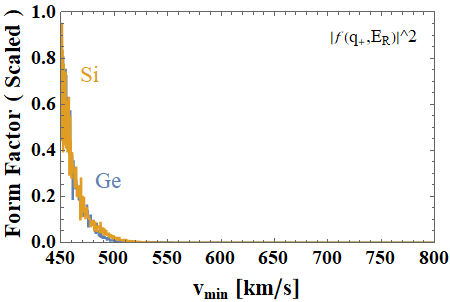}
\includegraphics[trim={0mm 0mm 0 0},clip,width=.49\textwidth]{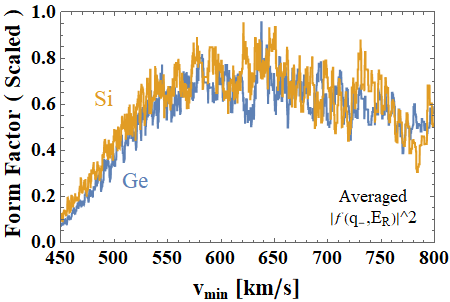}
\includegraphics[trim={0mm 0mm 0 0},clip,width=.49\textwidth]{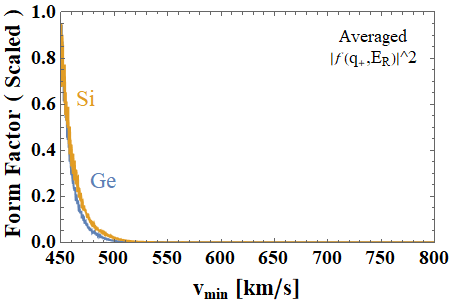}
\includegraphics[trim={0mm 0mm 0 0},clip,width=.49\textwidth]{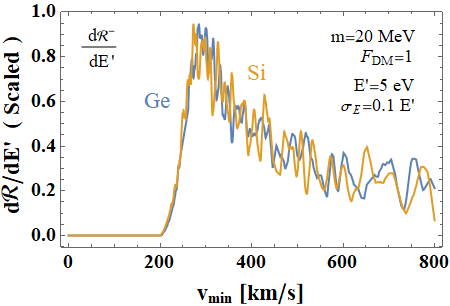}
\includegraphics[trim={0mm 0mm 0 0},clip,width=.49\textwidth]{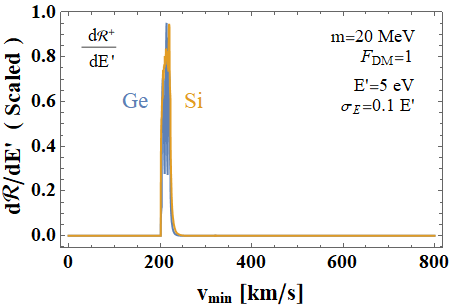}
\caption{ \label{fig:SiGe} 
Comparison of Ge and Si electron form factors $|f_{\rm crys}(q_{\pm}(\vmin,\ER),\ER)|^2$ in \Eq{eq:Gekernel} (two upper rows) for $\ER=10$ eV (top row) and average form factor in the $\ER$ range 9 eV to 11 eV (middle row),  and  Ge and Si response functions ${\ud \mathcal{R}^{\pm}_{\rm crys}(\vmin,E')}/{\ud E'}$ in \Eq{eq:Gekernel} for $m=$ 20 MeV, $E'=$ 5 eV, $\sigma_E=0.5$ eV and $F_{\rm DM}= 1$ (bottom row), as function of $\vmin$ for both the $q_{-}$ branch (left column) and the  $q_{+}$ branch (right column).
}
\end{center} 
\end{figure*}


\begin{figure*}[tbhp]
\begin{center}
\includegraphics[trim={0mm 18.5mm 0 0},clip,width=.49\textwidth]{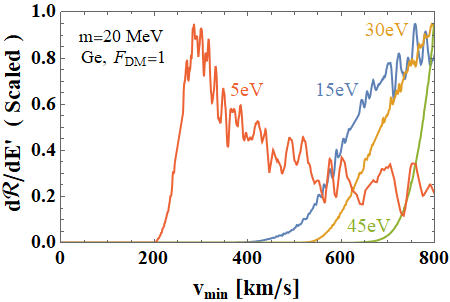}
\includegraphics[trim={0mm 18.5mm 0 0},clip,width=.49\textwidth]{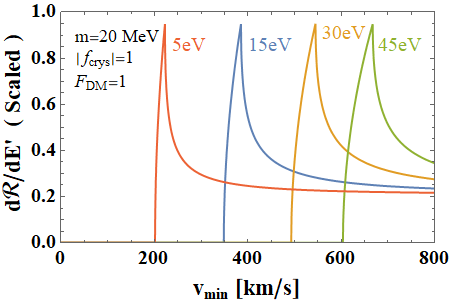} 
\includegraphics[trim={0mm 18.5mm 0 0},clip,width=.49\textwidth]{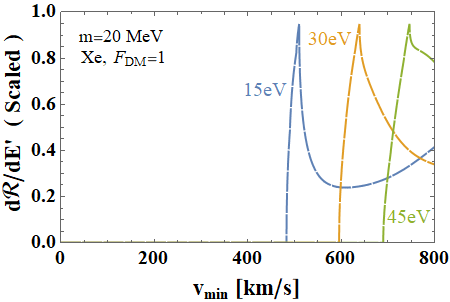} 
\includegraphics[trim={0mm 18.5mm 0 0},clip,width=.49\textwidth]{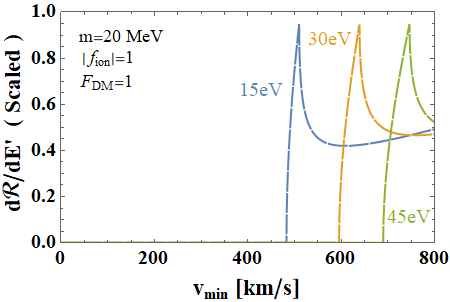}
\includegraphics[trim={0mm 18.5mm 0 0},clip,width=.49\textwidth]{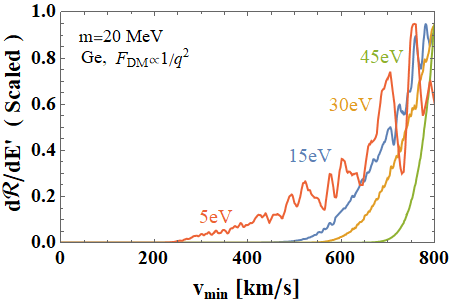}
\includegraphics[trim={0mm 18.5mm 0 0},clip,width=.49\textwidth]{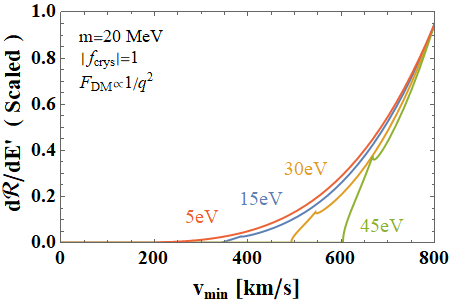}
\includegraphics[trim={0mm 0mm 0 0},clip,width=.49\textwidth]{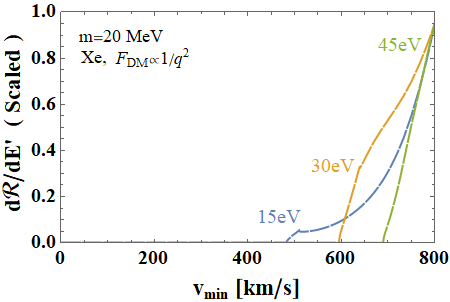} 
\includegraphics[trim={0mm 0mm 0 0},clip,width=.49\textwidth]{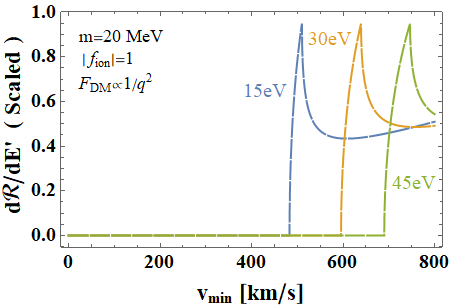}
\caption{Response functions $d\mathcal{R} (\vmin,E')/dE'$  for the time-average of $\tilde{\eta}(\vmin, t)$ in \Eq{eq:tilde-eta-t} scaled by their maxima (so each ratio has a maximum $\simeq$ 1) for energies $E'=$ 15, 30, and 45 eV and additionally $E'=$5 eV for Ge (in Xe, the minimum detectable energy is 13.8 eV), for  Ge (rows 1 and 3) and Xe detectors (rows 2 and 4), with actual electron form factors $|f^{i,f}|$ (left panels) and $|f^{i,f}|=1$ (right panels), $F_{\rm DM}=1$ (upper panels) and $\sim 1/q^2$ (lower panels), and $m=$20 MeV.  Halo properties can be inferred from data only where $d\mathcal{R}/ dE'\neq 0$. } 
\label{fig:20MeV}
\end{center}
\end{figure*}
\FloatBarrier


\begin{figure*}[tb]
\begin{center}
\includegraphics[trim={0mm 18.5mm 0 0},clip,width=.49\textwidth]{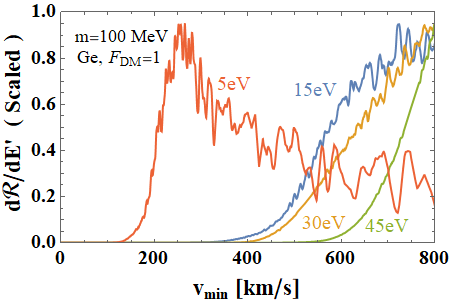} 
\includegraphics[trim={0mm 18.5mm 0 0},clip,width=.49\textwidth]{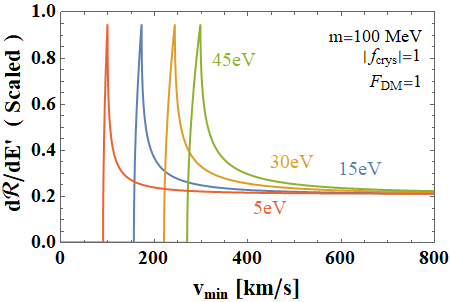} 
\includegraphics[trim={0mm 18.5mm 0 0},clip,width=.49\textwidth]{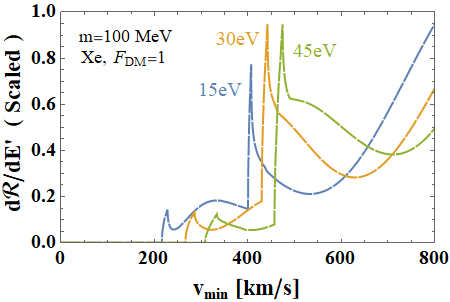} 
\includegraphics[trim={0mm 18.5mm 0 0},clip,width=.49\textwidth]{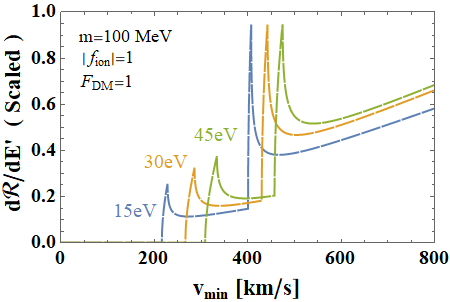}
\includegraphics[trim={0mm 18.5mm 0 0},clip,width=.49\textwidth]{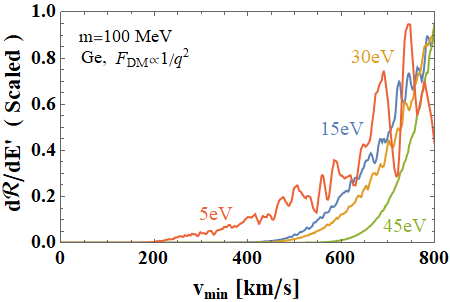}
\includegraphics[trim={0mm 18.5mm 0 0},clip,width=.49\textwidth]{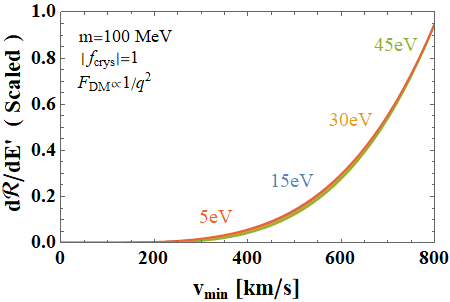} 
\includegraphics[trim={0mm 0mm 0 0},clip,width=.49\textwidth]{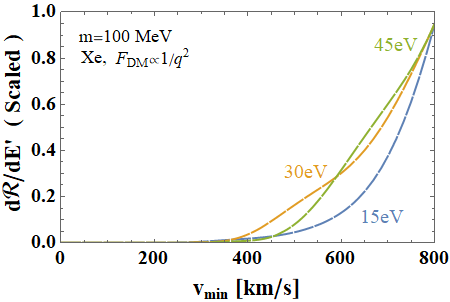} 
\includegraphics[trim={0mm 0mm 0 0},clip,width=.49\textwidth]{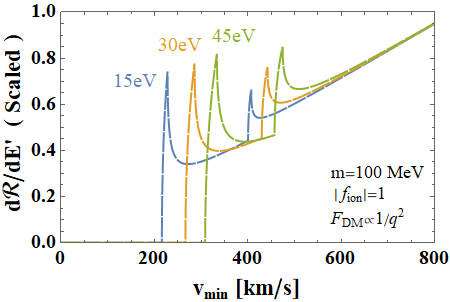}
\caption{ \label{fig:100MeV} 
Same as Fig~\ref{fig:20MeV}, but for a DM particle mass  $m=$ 100 MeV.
}
\end{center}
\end{figure*}

\FloatBarrier


\begin{figure*}[tb]
\begin{center}
\includegraphics[trim={0mm 18.5mm 0 0},clip,width=.49\textwidth]{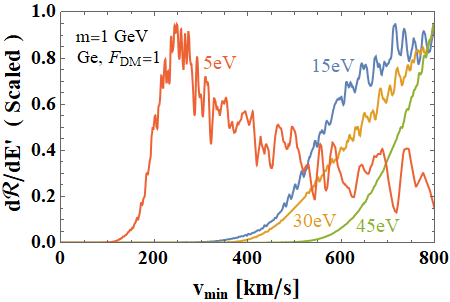} 
\includegraphics[trim={0mm 18.5mm 0 0},clip,width=.49\textwidth]{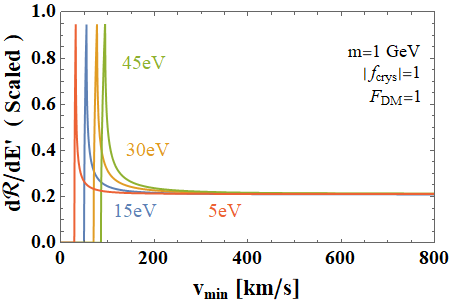} 
\includegraphics[trim={0mm 18.5mm 0 0},clip,width=.49\textwidth]{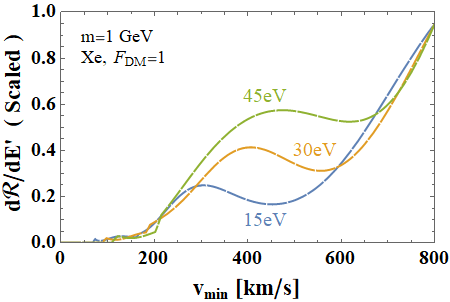} 
\includegraphics[trim={0mm 18.5mm 0 0},clip,width=.49\textwidth]{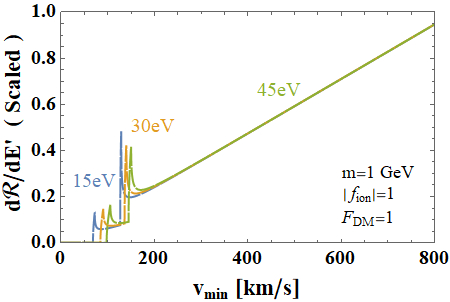}
\includegraphics[trim={0mm 18.5mm 0 0},clip,width=.49\textwidth]{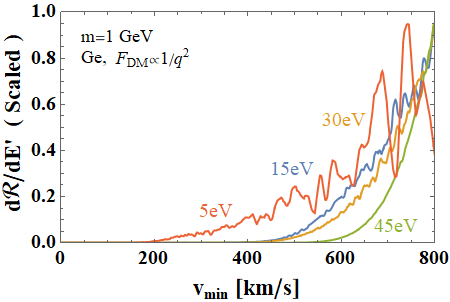}
\includegraphics[trim={0mm 18.5mm 0 0},clip,width=.49\textwidth]{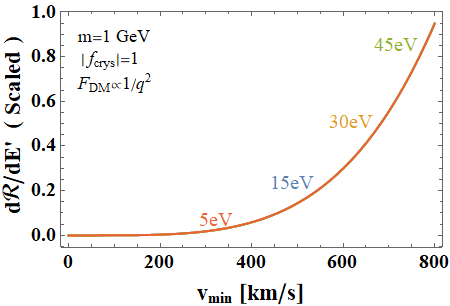} 
\includegraphics[trim={0mm 0mm 0 0},clip,width=.49\textwidth]{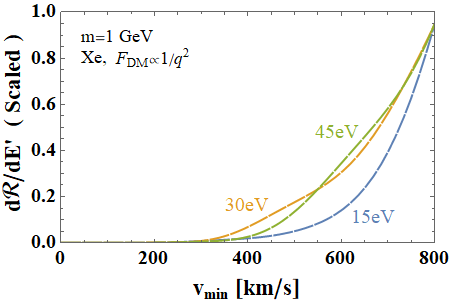} 
\includegraphics[trim={0mm 0mm 0 0},clip,width=.49\textwidth]{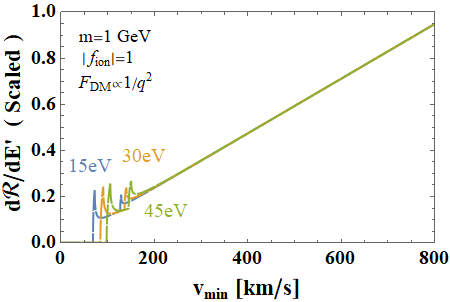}
\caption{ \label{fig:1000MeV} 
Same as Fig~\ref{fig:20MeV}, but for a DM particle mass  $m=$ 1 GeV.
}
\end{center}
\end{figure*}

\FloatBarrier

\begin{figure*}[t]
\begin{center}
\includegraphics[trim={0mm 18.5mm 0 0},clip,width=.49\textwidth]{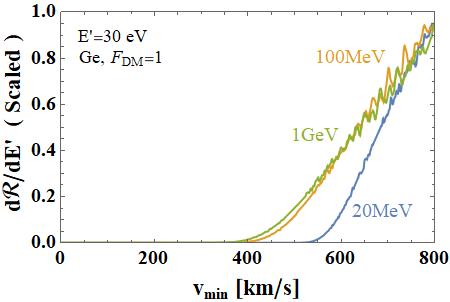} 
\includegraphics[trim={0mm 18.5mm 0 0},clip,width=.49\textwidth]{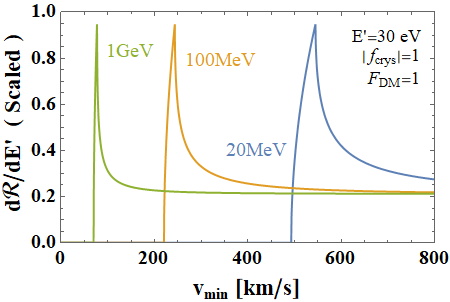} 
\includegraphics[trim={0mm 18.5mm 0 0},clip,width=.49\textwidth]{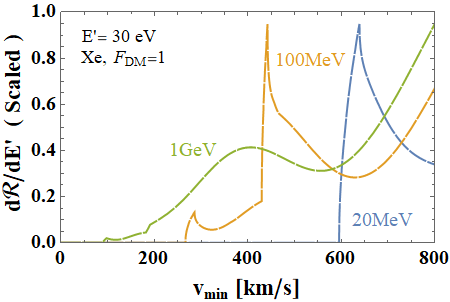} 
\includegraphics[trim={0mm 18.5mm 0 0},clip,width=.49\textwidth]{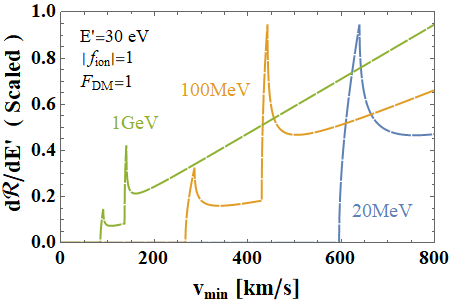}
\caption{ \label{fig:mass-comp} 
Response functions $d\mathcal{R} (\vmin,E')/dE'$  for the time-average of $\tilde{\eta}(\vmin, t)$ in \Eq{eq:tilde-eta-t} scaled by their maxima (so each ratio has a maximum $\simeq$ 1) for $F_{\rm DM}=1$ and energy $E'=$ 30 eV, $\sigma_E=$ 3.0 eV and DM mass of 20 MeV, 100 MeV and 1 GeV,  for Ge (row 1) and Xe detectors (rows 2), with the  actual electron form factors $|f^{i,f}|$ in the left panels and $|f^{i,f}|=1$ in the right panels.
}
\end{center}
\end{figure*}


\section{Results and Discussion} 
 \label{sec:resultdisc}

As we can see  in~\Fig{fig:SiGe}, Si and Ge crystals have electron form factors $|f_{\rm crys}(q_{\pm},\ER)|^2$ in \Eq{eq:Gekernel}  (shown in the upper panels of~\Fig{fig:SiGe}) and response functions (shown in the bottom row) of similar shape. Thus, for our purpose it is sufficient to present results only for one of the two semiconductor materials, and we arbitrarily chose Ge. 

Notice that in~\Fig{fig:SiGe}, as well as in the subsequent figures, Figs.~\ref{fig:20MeV} to \ref{fig:mass-comp}, we are interested in showing the range of $\vmin$ where the response functions are different from zero, i.e. the range of $\vmin$ for which a particular experiment measuring events at a particular energy can give information about the DM halo local velocity distribution. Recall that the response function acts as a window function through which measured rates in direct detection experiments can provide information about the DM velocity (or speed) distribution.  Since we are interested in the shape and not the magnitude of the functions, we show all of them divided by their maximum value,  hence in our plots we show  these dimensionless ratios with maximum value close to 1.

Figs.~\ref{fig:20MeV},~\ref{fig:100MeV} and ~\ref{fig:1000MeV}  display the response functions $\ud \mathcal{R}/ \ud E'$ for the time-average of $\tilde\eta(\vmin, t)$ in \Eq{eq:tilde-eta-t} as functions of $\vmin$, in Xe and Ge detectors, for DM particle masses of 20 MeV, 100 MeV and 1 GeV respectively and for two different DM form factors. We show these functions up to a maximum $\vmin$ value of $v_{\rm max}=$ 800 km/s, however our formalism can be trivially extended to larger  $v_{\rm max}$ to account for a possible contribution of DM not bound to the Galaxy.
These figures show the response functions  for Ge (rows 1 and 3) and Xe detectors (rows 2 and 4) assuming realistic detected energy values  $E'=$15, 30 and 45 eV and additionally $E'=$5 eV for Ge (in Xe, the minimum detectable energy is 13.8 eV). The four upper panels are for a DM form factor $F_{\rm DM} =1$, and the four lower panels assume $F_{\rm DM} \propto 1/q^2$. \Eq{Ipm-ion-sec3} and   \Eq{Ipm-crystal-sec3} show the effect of these different DM form factors in the $\vmin$ dependence of the response functions.  The actual electron form factors $|f^{i,f}|$ are used in the left panels, but the electron form factors are set to 1,  $|f^{i,f}|=1$, in the right columns.
Thus the comparison of the left and right panels of every row allows to easily understand the impact the actual electron form factors have on the shape of the response functions. The electron form factors significantly affect the response functions shape, moving their maxima and thresholds to larger $\vmin$ values. In semiconductor crystals, they also  introduce significant small scale variability in the function.

The double peaked structure of the Xe response functions seen in the right panels of  Figs.~\ref{fig:100MeV} and ~\ref{fig:1000MeV} is due to the inclusion of only the  4d and 5p initial electron orbitals in our numerical evaluations. These are the two orbitals that contribute dominantly to the response functions (as shown in Fig.~\ref{fig:Orb}). The double peaked form in not present in Fig.~\ref{fig:20MeV} because for $m=20$ MeV only the 4d orbital contributes for $\vmin < 800$ km/s.  

In all our figures we chose to parameterize the experimental energy resolution as a box function of width $2 \sigma_{\rm E}$ centered at the measured energy $E'$. In Figs.~\ref{fig:20MeV} to \ref{fig:mass-comp} we chose   $\sigma_{\rm E}= 0.10 E'$. Similar results are obtained with other forms for this function, such as a Gaussian.

With our choice of energy resolution, the minimum $\vmin$ value for which the resolution function is non-zero in each case is $\tilde{v}=\sqrt{2 (E' - \sigma_{\rm E}+ E_{\rm B})/m}$ (the function $\tilde{v}(E_e)$ is defined in \Eq{def-vtilde}), which decreases as $E'$ decreases and as $m$ increases. Fig.~\ref{fig:mass-comp} shows how the $\vmin$ threshold decreases with increasing DM particle mass, for a fixed $E'=30$ eV in Ge and Xe. The kinematic effect of $E_{\rm B} \neq 0$ is clearly seen in the right panels of Fig.~\ref{fig:mass-comp}, where the electron form factors are taken to be 1. When the actual form factors are considered they affect considerably the $\vmin$ thresholds (the minimum value of $\vmin$ for which a response function is significantly different from zero).
Only for DM masses close to 10 MeV (see our plots for $m= 20$ MeV), due to the initial binding energy $E_{\rm B}$ in Xe, for the same DM particle model and $E'$  the $\vmin$ threshold is lower in Ge and Si detectors than in Xe detectors. Instead, for larger masses (e.g. $m= 100$ MeV and 1 GeV) the electron form factors make the $\vmin$ threshold  lower in Xe than in  Ge and Si for the same DM mass and $E'$, although the effect is much less pronounced for $F_{\rm DM} \sim 1/ q^2$ than for $F_{\rm DM} = 1$.
However, $E'$ itself can be lower in semiconductor detectors, leading always to a smaller $\vmin$ threshold  in Ge and Si than in Xe for the same particle model.

Thus, in general, Si and Ge detectors extend to lower $\vmin$ values, than Xe detectors. However, this advantage of semiconductor based detectors diminishes for larger DM mass values. As an example,  see in the top panels of Fig.~\ref{fig:1000MeV} that for $m=1$ GeV the $\vmin$ threshold for $E'=5$ eV in Ge and that for $E'>15$ eV in Xe are both about 100 km/s in the left panels.

The response functions act also as a weight, indicating the range of $\vmin$ from which potential DM signal mostly originates. The most likely speeds of the DM producing the observed rate are those for which the response function is larger. Conversely, therefore, the DM halo function is more precisely determined in the $\vmin$ regions where the response function is larger. The figures show that for $F_{\rm DM} \propto 1/q^2$ the response functions weigh more the high speed tail of the halo function. 
Although here we considered only elastic scattering off electrons our results can be easily transferred to the case of inelastic scattering, as explained in Sec.~\ref{ssec:inelastic}.

 
  \section{Conclusions}
\label{sec:conclusion}
 
 We have shown how to extend to DM scattering off electrons the halo-independent direct detection data analysis, so far developed for and only applied to
 DM scattering off nuclei. 
 
A halo-independent analysis relies on the separation of the astrophysical parameters from the particle physics and detector dependent quantities. Namely, the method relies on expressing the predicted rate as a convolution of a function which depends only on the local DM distribution and density, and is thus common to all direct detection experiments, and a detector and DM particle model dependent kernel, which we call response function. The latter is non zero only for a limited range of DM particle speed (when considering just time-average rates and  isotropic interactions, otherwise the response function depends on the DM velocity vector), which depends on the detected energy. Thus, the response function acts as a window through which direct detection  data  can give information on the local dark halo.
Particular detectors can only provide information on the DM velocity distribution in a limited range of speeds depending on their characteristics. Thus complementarity of different experiments sensitive to partially overlapping velocity ranges would be needed to enable a more detailed inference of the DM distribution.
 
 We showed here the general definitions of the response functions for DM scattering off electrons for time-averaged rates and discussed the differences and similarities with the case of scattering off nuclei. As illustrations of the procedure, using the electron form factors provided in Ref.~\cite{Essig:2015cda}, we computed numerically the response functions for ionization of xenon atoms and excitation of electrons in silicon and germanium crystals. We  showed the range of DM speeds that experiments using 
 these materials can be sensitive to, for different
 DM particle masses, and for realistic detected energies and experimental energy resolutions. In general, experiments using semiconductor targets can have smaller energy thresholds and thus reach smaller DM speeds, while xenon based experiments due to their larger exposure would allow for a more precise determination of the halo properties in the speed range where they are sensitive. Thus these two types of experiments are complementary.  
 
 The particular response functions we computed are those which correspond to the time-average of the halo function $\tilde\eta(\vmin, t)$ (see \Eq{eq:tilde-eta-t}). However, in the 
 halo-independent method any integral of the DM speed (or velocity, in more complicated situations)
 can be used to characterize the halo, taking into account that any likelihood used to fit a halo function to direct detection data can be maximized by a DM speed  (or velocity)
 distribution written as a linear combination of delta functions, with a maximum number of terms given by the number of data points.
 
\acknowledgments
\addcontentsline{toc}{section}{Acknowledgments}

We thank Paolo Gondolo for participating in initial stages of this work. The work of G.B.G., P.L. and V.T. was supported in part by the U.S. Department of Energy (DOE) Grant No. DE-SC0009937. V.T. was also supported by World Premier International Research Center Initiative (WPI), MEXT, Japan.

\appendix

\section{Derivation of the general response function}
\label{Appendix-A}

The event rate of collisions in which an electron jumps from an initial state $i$ to a final state $f$ with energy difference $\Delta E_{i\rightarrow f}$ equal to the energy $E_e$ lost by the DM particle  is
\begin{eqnarray}
 R_{i\rightarrow f}=\frac{\rho}{m}\int \ud^3 v\, f_\chi (\vec{v})\langle \sigma v\rangle_{i\rightarrow f}~,
\end{eqnarray}
where,
\begin{eqnarray}
 \langle \sigma v\rangle_{i \rightarrow f}=\frac{\overline{\sigma}_e}{4\pi\mu^2_{\chi e}} \int \ud^3 q\, \delta(\Delta E_{i\rightarrow f}-\frac{q^2}{2m}+qv\cos \theta_{qv})|F_{\rm DM}(q)|^2|f_{i\rightarrow f}(\vec{q})|^2~.
\end{eqnarray}
Not all of the energy $E_e$ given to the electron in the collision is detectable. We call $\ER$ the detectable energy, and  
\begin{eqnarray}
\ER=\Delta E_{i\rightarrow f}-E_{{\rm B}i}~.
\end{eqnarray}
$\ER$ is the electron recoil energy if electrons are free in the final state, e.g. electrons in an atom that is ionized, for which $E_{{\rm B}i}$ is the initial binding energy. Instead, neglecting very small energy losses, such as thermal dissipation via phonons, we take $\ER= E_e$ for electrons scattering within a crystal. This condition can be written as a delta function 
\begin{equation}
1=\int \ud \ER\,  \delta(\ER-\Delta E_{i \rightarrow f}+E_{{\rm B}i})~,
\end{equation}
which is incorporated in the particular transition rate,
\begin{eqnarray}
R_{i\rightarrow f}&=&\frac{\rho}{m}\frac{\overline{\sigma}_e}{4\pi \mu^2_{\chi e}}\int \ud^3v\,\ud^3q\,\ud \ER\, \delta(\ER+ E_{{\rm B}i}-\frac{q^2}{2m}+qv\cos\theta_{qv})f_\chi(\vec{v})\nonumber \\
&&\times \delta(\ER-\Delta E_{i\rightarrow f}+E_{{\rm B}i})|F_{\rm DM}(q)|^2|f_{i\rightarrow f}(\vec{q})|^2~.
\end{eqnarray}
Then, summing over all possible occupied initial states and final states, the total rate is 
\begin{eqnarray} \label{rate-summed}
R&=& \sum_{i,f} 
\frac{\rho}{m}\frac{\overline{\sigma}_e}{4\pi \mu^2_{\chi e}}\int \ud^3v\,\ud^3q\,\ud\ln\ER\, \delta(\ER+E_{{\rm B}i}-\frac{q^2}{2m}+qv\cos\theta_{qv})f_\chi(\vec{v})\nonumber \\
&&\times \ER\delta(\ER-\Delta E_{i\rightarrow f}+E_{{\rm B}i})|F_{\rm DM}(q)|^2|f_{i\rightarrow f}(\vec{q})|^2~.
\end{eqnarray}
Here the summations represent integrations over continuous quantum numbers and actual summations over the discrete ones.  The summation can be separated into sums over energy levels (indicated with a prime) and sums over all degenerate states,
\begin{eqnarray}
\label{eq:sumif}
\sum_i \sum_f = \sideset{}{'}\sum_{i,f} \times \sum_{\substack{\rm degen.\\ \rm states}}~.
\end{eqnarray}
 The sum of the electron form factors $|f_{i\rightarrow f}(\vec{q})|^2$ over degenerate states which we call $|f^{i,f}|^2$, includes summing over all directions. Thus, there is no angular dependence left in $|f^{i,f}|^2$. 
Therefore, the sum can depend only on $q= |\vec{q}|$ (this is \Eq{fij-def-Sec3})
\begin{eqnarray} \label{fif-definition}   
|f^{i,f}(q,\ER)|^2= \sum_{\substack{ \rm degen.\\ \rm states}} \ER~\delta(\ER-\Delta E_{i\rightarrow f}+E_{{\rm B}i})|f_{i\rightarrow f}(\vec{q})|^2~.
\end{eqnarray}
Using the following relation 
\begin{equation} \label{delta-cos}
\delta(\ER+E_{{\rm B}i}-\frac{q^2}{2m}+qv\cos\theta_{qv}) = \frac{1}{qv} \delta(\cos\theta_{qv}-\frac{\vmin}{v})~,
\end{equation}
we can perform the  momentum integral over the solid angle $\Omega_{qv}$ in the rate to get
\begin{eqnarray}
R&=& \sideset{}{'}\sum_{i,f}  \frac{1}{2 \mu^2_{\chi e}}\int \ud\ln\ER\int \ud q\, |F_{\rm DM}(q)|^2|f^{i,f}(q,\ER)|^2q \nonumber\\
&&\times \left\{ \frac{\rho \overline{\sigma}_e}{m}\int \ud^3v \, \frac{1}{v}\Theta(v-\vmin(q,\ER+E_{{\rm B}i}))f_\chi(\vec{v}) \right\}~.
\end{eqnarray}
The term inside the curly brackets in this equation  is $\tilde{\eta}(\vmin(q,\ER+E_{{\rm B}i} ))$,  the time-average of the function defined in \Eq{eq:tilde-eta-t} with the reference cross section $\sigma_{\rm ref}= \overline{\sigma}_e$.
Thus, the general equation for the differential rate is
\begin{eqnarray} 
\label{eq: App-A-general-dRdER}
\frac{\ud R}{\ud\ER}=\frac{1}{2\mu^2_{\chi e}}\frac{1}{\ER}\sideset{}{'}\sum_{i,f} \int \ud q\,\tilde{\eta}(\vmin(q,\ER+E_{{\rm B}i})) |F_{\rm DM}(q)|^2|f^{i,f}(q,\ER)|^2q~.
\end{eqnarray}

\section{Electron form factor for DM scattering off electrons in an atom}
\label{app:formion}

In an atom target, the initial bound states are labeled by the principal quantum number $n$ and  the angular momentum  quantum number $l$, $|i\rangle=| nl \rangle$. If after the collision the atom is ionized, the final states are free spherical wave states labeled by $|f\rangle=| k'l'm'\rangle$, where $k'$ is the magnitude of the free electron momentum and $m', l'$ are spherical harmonic labels ($l'$, $m'$ labeled states are degenerate for the same $k'$). Thus, 
 the form factor $|f_{i\to f}(\vec{q})|$ in \Eq{rate-summed} is now
\begin{equation}
  |f_{i\rightarrow f}(\vec{q})|^2=  |f_{nl\rightarrow k'l'm'}(\vec{q})|^2
  =\left|\int \ud^3 x \, \psi_{k'l'm'}^*(\vec{x})\psi_{nl}(\vec{x})e^{i\vec{q}\cdot\vec{x}}\right|^2 ~,
\end{equation}
and the summation over states given in \Eq{eq:sumif} becomes
\begin{eqnarray} 
\sum_{i,f}=\sum_{nl} \sum_{\substack{ \rm degen.\\ \rm states}} \int \frac{k'^2\ud k'}{(2\pi)^3}~.
\end{eqnarray}
Summed over all final states, the form factor given in \Eq{fif-definition} becomes
\begin{eqnarray}  \label{fij-atom-2}           
\int \frac{k'^2 \ud k'}{(2\pi)^3}\,|f^{nl, k'}_{\rm ion}(q,\ER)|^2 
&=&  \sum_{\substack{ \rm degen.\\ \rm states}} \int \frac{k'^2 \ud k'}{(2\pi)^3}\,\ER\delta(\ER-\Delta E_{nl\rightarrow k'l'm'}+E_{{\rm B}nl})\\
&&\times\left|\int \ud^3 x \, \psi_{k'l'm'}^*(\vec{x})\psi_{nl}(\vec{x})e^{i\vec{q}\cdot\vec{x}}\right|^2 \notag
\end{eqnarray}
Since the recoil energy is 
\begin{eqnarray}
\ER'=\frac{k'^2}{2m_e}=\Delta E_{nl\rightarrow k'l'm'}-E_{{\rm B}nl}~,
\end{eqnarray}
we have
\begin{eqnarray}  \label{fij-atom}          
\int \frac{k'^2 \ud k'}{(2\pi)^3}&&|f^{nl, k'}_{\rm ion}(q,\ER)|^2 
= \sum_{\substack{ \rm degen.\\ \rm states}} \int \frac{k'^3 \ud\ln\ER'}{2(2\pi)^3}\,\ER\delta(\ER-\ER')\left|\int \ud^3 x \, \psi_{k'l'm'}^*(\vec{x})\psi_{nl}(\vec{x})e^{i\vec{q}\cdot\vec{x}}\right|^2 \nonumber\\
&=& \sum_{\substack{ \rm degen.\\ \rm states}}\, \frac{k'^3}{2(2\pi)^3}\left|\int \ud^3 x \, \psi_{k'l'm'}^*(\vec{x})\psi_{nl}(\vec{x})e^{i\vec{q}\cdot\vec{x}}\right|^2
=\dfrac{1}{4}|f_{\rm ion}^{nl}(q,\ER)|^2~.
\end{eqnarray}
Where we have used the definition of the form factor  given in  Eq.(6) of Ref.~\cite{Essig:2011nj},
\begin{eqnarray}
|f_{\rm ion}^{nl}(q,\ER)|^2=\sum_{\substack{ \rm degen.\\ \rm states}}\, \frac{2k'^3}{(2\pi)^3}\left|\int \ud^3 x \, \psi_{k'l'm'}^*(\vec{x})\psi_{nl}(\vec{x})e^{i\vec{q}\cdot\vec{x}}\right|^2.
\end{eqnarray}

\section{Electron form factor for DM scattering off electrons in a crystals}
\label{app:formcrystal}

In crystals, a DM particle excites an electron  from a Bloch state $|{n\vec{k}}\rangle$ to another Bloch state $|{n'\vec{k}'}\rangle$, where $n$ and $n'$ are the band index labels in the first Brillouin Zone (BZ).
Thus,  the form factor $|f_{i\to f}(\vec{q})|$ in  \Eq{rate-summed} is
\begin{equation}
  |f_{i\rightarrow f}(\vec{q})|^2=  |f_{n\vec{k}\rightarrow n'\vec{k}'}(\vec{q})|^2
  =\left|\int \ud^3 x \, \psi_{n'\vec{k}'}^*(\vec{x})\psi_{n\vec{k}}(\vec{x})e^{i\vec{q}\cdot\vec{x}}\right|^2 ~.
\end{equation}
The Bloch state wavefunctions $\psi_{n\vec{k}}$ are
\begin{eqnarray}
\label{eq:blochstate}
\psi_{n\vec{k}}(\vec{x})=\dfrac{1}{\sqrt{V}}\sum_{\vec{G}}u_{n}(\vec{k}+\vec{G})e^{i(\vec{k}+\vec{G})\cdot\vec{x}}~,
\end{eqnarray}
where the summation is over all reciprocal lattice vectors $\vec{G}$, $V$ is the volume of the crystal, and $u_n$ are normalized wavefunctions satisfying
\begin{equation}
    \sum_{\vec{G}}|u_{n}(\vec{k}+\vec{G})|^2=1~.
\end{equation}
Thus,
\begin{eqnarray} \label{Bloch-form-factor}
|f_{n\vec{k}\rightarrow n'\vec{k}'}(\vec{q})|^2 
=\left| \int \ud^3 x\, \frac{1}{V}\sum_{\vec{G}'}u^*_{n'}(\vec{k}'+\vec{G}')e^{i(\vec{k}'+\vec{G}')\cdot \vec{x}} \sum_{\vec{G}}u_{n}(\vec{k}+\vec{G})e^{i(\vec{k}+\vec{G})\cdot \vec{x}} e^{i\vec{q}\cdot\vec{x}}~\right|^2~.
\end{eqnarray}
 For crystal dimensions much larger than the atomic separation (we consider only large detectors where the interactions happen in the volume and surface effects are negligible) the discrete crystal momentum $ \vec{k}$ can be approximated by a continuous variable, thus 
\begin{eqnarray}
\label{eq:del3k}
\int \ud^3 x ~ e^{i \vec{k}\cdot\vec{x}}=(2\pi)^3\delta^3(\vec{k})~,
\end{eqnarray}
and the summation over states in \Eq{eq:sumif} becomes
\begin{eqnarray} \label{total-sum-crystal}
\sum_{i,f}=2\sum_{nn'} \int_{\rm BZ} \frac{V\ud^3 k}{(2\pi)^3}\frac{V\ud^3 k'}{(2\pi)^3}~.
\end{eqnarray}
Here, the extra factor of 2 comes from summing over degenerate spin states.
Since the summation in \Eq{eq:blochstate} is over all reciprocal vectors, 
\begin{eqnarray}
\frac{1}{\sqrt{V}}\sum_{\vec{G}'} u_n(\vec{k}'+\vec{G}')e^{i(\vec{k}'+\vec{G}')\cdot\vec{x}}=\frac{1}{\sqrt{V}}
\sum_{\vec{G}'}u_n(\vec{k}'+\vec{G}'+\vec{G})e^{i(\vec{k}'+\vec{G}'+\vec{G})\cdot\vec{x}}~,
\end{eqnarray}
 using \Eq{eq:del3k}, one can write the form factor in \Eq{Bloch-form-factor} as
\begin{eqnarray} 
|f_{n\vec{k}\rightarrow n'\vec{k}'}(\vec{q})|^2=\left| \sum_{\vec{G}}\sum_{\vec{G}'} \frac{(2\pi)^3}{V}\delta^3(\vec{k}+\vec{q}-\vec{k}'-\vec{G}') u^*_{n'}(\vec{k}'+\vec{G}'+\vec{G})u_{n}(\vec{k}+\vec{G})\right|^2~.
\end{eqnarray}
Since all other terms in~\Eq{eq: App-A-general-dRdER} besides the electron form factors are independent of the choice of $i$ and $f$ states,  as a result of taking $E_{{\rm B}i}=0$, the summation over initial and final states with different energy only affects the electron form factors, and  this sum is

\begin{equation}  \label{intermediate-crystal}
\sideset{}{'}\sum_{i,f}|f^{i,f}(q,\ER)|^2=2\sum_{nn'} \int_{\rm BZ} \frac{V\ud^3 k}{(2\pi)^3}\frac{V\ud^3 k'}{(2\pi)^3}\,\ER \delta(\ER-\Delta E_{n\vec{k}\rightarrow n'\vec{k}'})|f_{n\vec{k}\rightarrow n'\vec{k}'}(\vec{q})|^2~.
\end{equation}

The summation in \Eq{intermediate-crystal} includes a summation over directions, thus the result is independent of the direction of $\vec{q}$, a result which can be incorporated  into \Eq{intermediate-crystal} by replacing in it $|f_{n\vec{k}\rightarrow n'\vec{k}'}(\vec{q})|^2$ by
\begin{eqnarray}
\int \ud^3 p\,\delta^3(\vec{p}-\vec{q})   |f_{n\vec{k}\rightarrow n'\vec{k}'}(\vec{p})|^2=\int \dfrac{\ud^3p}{4\pi p^2}\delta(p-q)  |f_{n\vec{k}\rightarrow n'\vec{k}'}(p)|^2~.
\end{eqnarray}
\Eq{intermediate-crystal} becomes
\begin{eqnarray}  
\sideset{}{'}\sum_{i,f}|f^{i,f}(q,\ER)|^2&=&
\int \dfrac{\ud^3p}{2\pi p^2}\,\delta(p-q)\sum_{nn'} \int_{\rm BZ} \frac{V^2\ud^3 k \,\ud^3 k'}{(2\pi)^6}\,\ER \delta(\ER-\Delta E_{n\vec{k}\rightarrow n'\vec{k}'}) |f_{n\vec{k}\rightarrow n'\vec{k}'}(p)|^2 \nonumber\\
&=&\frac{4\pi^2}{ q^3 V} \sum_{nn'} \int_{\rm BZ} \frac{V\ud^3 k }{(2\pi)^3}\frac{V\ud^3 k'}{(2\pi)^3}\,\ER \delta(\ER-\Delta E_{n\vec{k}\rightarrow n'\vec{k}'})\notag\\
&&\times\sum_{\vec{G}'} \,q\delta(q-|\vec{k}'+\vec{G}'-\vec{k}|)\left| \sum_{\vec{G}} u^*_{n'}(\vec{k}'+\vec{G}'+\vec{G})u_{n}(\vec{k}+\vec{G}) \right|^2.
\end{eqnarray}
Expressing the total volume in terms of the volume $V_{\rm cell}$ of one cell,  $V=N_{\rm cell}V_{\rm cell}$, 
\begin{eqnarray} \label{fij-crystal}
\sideset{}{'}\sum_{i,f}|f^{i,f}(q,\ER)|^2&=&\frac{4\pi^2N_{\rm cell}}{ q^3 V_{\rm cell}} \sum_{nn'} \int_{\rm BZ} \frac{V_{\rm cell}\,\ud^3 k }{(2\pi)^3}\frac{V_{\rm cell}\,\ud^3 k'}{(2\pi)^3}\,\ER \delta(\ER-\Delta E_{n\vec{k}\rightarrow n'\vec{k}'})\notag\\
&&\times\sum_{\vec{G}'} \,q\delta(q-|\vec{k}'+\vec{G}'-\vec{k}|)\left| \sum_{\vec{G}} u^*_{n'}(\vec{k}'+\vec{G}'+\vec{G})u_{n}(\vec{k}+\vec{G}) \right|^2\notag\\
&=&\frac{2N_{\rm cell}\ER}{q^3}(\alpha m_e^2)|f_{\rm crys}(q,\ER)|^2~,
\end{eqnarray}
where 
\begin{eqnarray}
|f_{\rm crys}(q,\ER)|^2&=&\frac{2\pi^2}{\alpha m_e^2}\frac{1}{V_{\rm cell}\ER}\sum_{nn'}\int_{\rm BZ} \frac{V_{\rm cell}\,\ud^3 k}{(2\pi)^3}\, \frac{V_{\rm cell}\,\ud^3 k'}{(2\pi)^3} \,\ER\delta(\ER-\Delta E_{n\vec{k}\rightarrow n'\vec{k}'})\nonumber\\
&&\times \sum_{\vec{G}'}q\delta(q-|\vec{k}'+\vec{G}'-\vec{k}|)\left| \sum_{\vec{G}} u^*_{n'}(\vec{k}'+\vec{G}'+\vec{G})u_{n}(\vec{k}+\vec{G}) \right|^2~.
\end{eqnarray}
is the crystal form factor defined in Eq.~(A.33) of Ref.~\cite{Essig:2015cda}.

\section{Expression for numerical calculations of the response function for general a, b and c constants}
\label{app:numres}

We use the computed electron form factors of Ref.~\cite{Essig:2015cda} for Xe, Ge and Si target, but our analysis is general and our treatment is readily applicable to any given electron form factors that can be described by binned data with a power law dependency of $q^a$ and $\ER^b$, where $a, b$ are integers, so that 
\begin{equation}
|f^{i,f }(q_{\pm},\ER)|^2=q_{\pm}^a\ER^b|f^{i,f}_{\rm dat}(\ER,q_{\pm})|^2~.
\end{equation}
In the particular cases we consider in this paper, $a=-3$ and $b=1$ for crystals (see Eq.~\eqref{eq:numrescrys}) and $a=b=0$ for an atom (see Eq.~\eqref{eq:numresion}). 
Writing additionally $|F_{\rm DM}(q)|^2= (\alpha m_e)^{2c}q^{-2c}$, where either $c=0$ or $c=2$ are our two choices, the response function integral to be evaluated in  Eq.~\eqref{eq:response-ER-int}  takes the functional form
\begin{eqnarray} \label{eq:dRint}
\frac{d\mathcal{R}_{\pm}}{dE'}(\vmin,E')&=& \frac{\epsilon(E')}{4\mu_{\chi e}^2 \sigma_{\rm E}}(\alpha m_e)^{2c}\sideset{}{'}\sum_{i,f}\int_{E'-\sigma_{\rm E}}^{E_{\rm max}} \ud\ER\,\ER^{b-1}~q^{a+1-2c}_{\pm}(\vmin,\ER+E_{{\rm B}i})~ \notag\\
& \times &J_{\pm}(\vmin,\ER+E_{{\rm B}i})
|f^{i,f}_{\rm dat}(\ER,q_{\pm}(\vmin,\ER+E_{{\rm B}i}))|^2~. 
\end{eqnarray}
Extracting the constant data value for the electron form factor in each bin and performing the remaining integral analytically,  the  expression we compute numerically is
\begin{eqnarray} \label{eq:genres}
\frac{d\mathcal{R_{\pm}}}{dE'}=\frac{\epsilon(E')}{4\mu_{\chi e}^2\sigma_{\rm E}}(\alpha m_e)^{2c}\sideset{}{'}\sum_{i,f}\sum_{{\rm Path}(r,s)} I^{\pm,i}_{abc}(\vmin, E_1^{(rs)},E_2^{(rs)})~|f^{i,f}_{\rm dat}|^2_{rs}~,
\end{eqnarray}
where the summation is over all the $(r,s)$ partitions through which the path described by the function $q(\vmin,\ER)$ passes, as indicated in \Fig{fig:partition}. The analytically computed $I^{\pm,i}_{abc}(E_1^{(rs)},E_2^{(rs)})$ integrals are
\begin{eqnarray}
\label{eq:Imn}
I^{\pm,i}_{abc}(\vmin, E_1^{(rs)},E_2^{(rs)})=\int_{E_1^{(rs)}}^{E_2^{(rs)}} \dfrac{\ud\ER}{\ER^{1-b}}\, q^{a+1-2c}_{\pm}(\vmin,\ER+E_{{\rm B}i}) J_{\pm}(\vmin,\ER+E_{{\rm B}i})~.
\end{eqnarray}
Here, $E_1^{(rs)}$ and $E_2^{(rs)}$ are the boundaries of the shaded area in the partition $(rs)$ shown in \Fig{fig:partition}. 
Since the partitions are small, we use the approximation $\ER\approx (E_1^{(rs)}+E_2^{(rs)})/2$ and take $\ER^{b-1}$ outside the integral, so \Eq{eq:Imn} becomes
\begin{equation}
I^{\pm,i}_{abc}(\vmin, E_1^{(rs)},E_2^{(rs)}) \simeq 2\int_{E_1^{(rs)}}^{E_2^{(rs)}}\ud\ER\,\frac{q^{a+1-2c}_{\pm}(\vmin,\ER+E_{{\rm B}i}) \,J_{\pm}(\vmin,\ER+E_{{\rm B}i})}{\left(E_1^{(rs)}+E_2^{(rs)}\right)^{1-b}} ~.
\end{equation}
Performing this integral we find:
for $a-2c=-3$,
\begin{eqnarray}
I^{\pm,i}_{abc}(\vmin, E_1^{(rs)},E_2^{(rs)}) \simeq
 \mp\left(\frac{E_1^{(rs)}+E_2^{(rs)}}{2}\right)^{b-1}\left.\left[\ln\left| 1\pm\sqrt{1-\dfrac{2(\ER+E_{{\rm B}i})}{m \vmin^2}} \right|\right]\right|_{E_1^{(rs)}}^{E_2^{(rs)}}~;
\end{eqnarray}
for $a-2c\neq -3$,
\begin{eqnarray}
I^{\pm,i}_{abc}(\vmin, E_1^{(rs)},E_2^{(rs)}) &\simeq&
 \mp\left(\frac{E_1^{(rs)}+E_2^{(rs)}}{2}\right)^{b-1}\dfrac{(m\vmin)^{a-2c+3}}{a-2c+3} \\
 &&\times\left.\left[\left( 1\pm\sqrt{1-\dfrac{2(\ER+E_{{\rm B}i})}{m\vmin^2}} \right)^{a-2c+3} \right]\right|_{E_1^{(rs)}}^{E_2^{(rs)}}\nonumber~.
\end{eqnarray}
Both of them are always positive.

\section{Alternative derivation of the response function for DM scattering off electrons} 
\label{app:curlyH} 

The response function  $\ud \mathcal{R}/\ud E'$ can be derived using the relation in \Eq{derivative-of-CurlyH}
\begin{equation} \label{derivative-of-CurlyH-App}
\frac{\ud \mathcal{R}}{\ud E'}(\vmin,E') \equiv \frac{\partial}{\partial \vmin}\left[ \frac{\ud \mathcal{H}}{\ud E'}(\vmin,E') \right] \, ,
\end{equation}
from the response function  ${\ud \mathcal{H}}(\vmin,E')/{\ud E'}$ to the speed distribution, when we write the rate as in \Eq{diffrate_manip1-nuc}, namely
\begin{equation}\label{drate-CurlyH-App}
 \frac{\ud R}{\ud E'} = \frac{\overline{\sigma}_e \rho}{m}\int_0^\infty{\rm d}v~\frac{F(v,t)}{v}\frac{{\rm d}\mathcal{H}}{{\rm d}E'} (v, E') \, .
 \end{equation}
 This procedure allows us to make contact with the formulation used for DM scattering off nuclei, in particular for inelastic scattering, as explained in App~\ref{app:nuclearscat}.

Starting from \Eq{rate-summed} and using \Eq{fif-definition} and \Eq{delta-cos} we get
\begin{eqnarray} \label{differential-rate-summed-App}
\dfrac{\ud R}{\ud \ER}= \sideset{}{'}\sum_{i,f} 
\frac{\rho~\overline{\sigma}_e}{4\pi \mu^2_{\chi e} m \ER}\int \ud^3v f_\chi(\vec{v}) \int \frac{\ud^3q}{qv} \delta(\cos\theta_{qv}-\frac{\vmin}{v}) |F_{\rm DM}(q)|^2|f^{i,f}(q,\ER)|^2,
\end{eqnarray}
or after further simplification
\begin{eqnarray} \label{differential-rate-speed-App}
\dfrac{\ud R}{\ud \ER}= \sideset{}{'}\sum_{i,f} 
\frac{\rho}{m}\frac{\overline{\sigma}_e}{2 \mu^2_{\chi e} \ER}
\int_{\tilde{v}} \ud v\, \frac{F(v)}{v} \int_{q_{-}(v,E_e)}^{q_{+}(v,E_e)} \ud q\, q~  
|F_{\rm DM}(q)|^2~|f^{i,f}(q,\ER)|^2~,
\end{eqnarray}
where $\tilde{v}=\sqrt{2 E_e/m}$ and  the functions $q_{\pm}(v,E_e)$ are given in \Eq{qbranches}. Integrating now on $\ER$ to obtain the observable differential rate  ${\ud R}/{\ud E'}$ we identify ${\ud \mathcal{H}}(E',v)/{\ud E'}$ as
\begin{eqnarray} \label{CurlyH-App}
\dfrac{\ud \mathcal{H}}{\ud E'} (v, E')  \equiv 
  \begin{dcases}  \hfill
\frac{\epsilon(E')}{2 \mu^2_{\chi e} } \int \frac{\ud \ER}{\ER} G(E',\ER)\sideset{}{'}\sum_{i,f}
 \int_{q_{-}(v,E_e)}^{q_{+}(v,E_e)} \ud q\, q  
|F_{\rm DM}(q)|^2|f^{i,f}(q,\ER)|^2  \hfill & \text{ if $v \geqslant \tilde{v}$,} \\
& \\
 \hfill 0 \hfill & \text{ if $v < \tilde{v}$.} \\
\end{dcases}
\end{eqnarray}
Notice that this expression depends on the speed $v$  only though the limits of the integration in $q$. Taking the partial derivative of this expression with respect to $v$, and recalling we defined in \Eq{Jacobian} ${\partial q_{\pm}(v,E_e)}/{\partial v}=J_{\pm}(v,E_e)$,  we recover Eqs.~\eqref{kernel-general} and \eqref{eq:response-ER-int}.

\section{Comparison with the formalism for DM scattering off nuclei}
\label{app:nuclearscat}

In DM-electron scattering, with kinematics described by Eq.~\eqref{vmin-of-q-Ee}, two of the three variables - momentum transfer $q$, detectable energy $\ER$, and  $\vmin$ - are independent. In the halo-dependent usual formalism, $\vmin$  is taken to be a function of the other two. In the halo-independent formalism, we choose instead $\ER$ and  $\vmin$ as independent variables, and thus we change
an integration in $q$ to obtain the rate into an integration in $\vmin$.

In the DM scattering off nuclei (of mass $m_N$) the kinematics sets $\ER=q^2/ 2 m_N$, and only one of the two variables $\ER$ and $\vmin$ is independent. In the halo-independent method we chose $\vmin$ to be the independent variable and thus change an integration in $\ER$ to get the predicted event rate into an integration in $\vmin$. 

In spite of these differences in the kinematics of DM-electron scattering and DM-nucleus  scattering, we can identify a similarity in the equations we get in $q$ in DM scattering off electrons and in $\ER$ in the inelastic endothermic DM scattering off nuclei. In this latter case, the DM particles scatter to a new state of mass $m^{\prime} = m + \delta$, where $|\delta| \ll m$, and $\delta > 0$ ($<0$) describes endothermic (exothermic) scattering. In this case (see e.g. Refs.~\cite{DelNobile:2013cva,Gelmini:2015voa,Gelmini:2016pei}), in the limit $\mu_{\chi N} |\delta|/m^2 \ll 1$, $\vmin(\ER)$ is
\begin{equation}\label{eq:vmin-delta}
\vmin(\ER) = \frac{1}{\sqrt{2 m_N \ER}} \left| \frac{m_N \ER}{\mu_{\chi N}} +\delta \right| \, ,
\end{equation}
which reduces to the typical equation for elastic scattering when $\delta = 0$.  Thus, the range of possible recoil energies that can be imparted to a target nucleus by a DM particle traveling at speed $v$ in Earth's frame is  $[\ER^{-}(v),\ER^{+}(v)]$, where
\begin{equation}\label{eq:Ebranch}
\ER^{\pm} (v) = \frac{\mu_{\chi N}^2 v^2}{2 m_N} \left( 1 \pm \sqrt{1-\frac{2\delta}{\mu_{\chi N} ~ v^2}} \right)^2 \, .
\end{equation}
For endothermic scattering the minimum possible value of $v$ is  $v_\delta^N = \sqrt{2 \delta / \mu_{\chi N}}> 0$,  which depends on the nuclide type through the reduced mass $\mu_{\chi N}$  (for exothermic and elastic scattering the minimum is instead $v_\delta =0$).
Notice the similarity of the definition of $\ER^{\pm} (v)$ in \Eq{eq:Ebranch} with the definition of $q_{\pm}(v, E_e)$  in \Eq{qbranches}. In \Eq{differential-rate-speed-App}, for a fixed speed $v$ the integration in $q$ is between $q_{-}$ and $q_{+}$, and similarly here the integration for fixed speed is between $\ER^{-}$ and $\ER^{+}$. The differential rate as a function of the detected energy $E'$ for DM scattering off a nuclide $N$ is (see \Eq{diffrate-tot}, \Eq{diffrate}, \Eq{eq:diffrate_withf} and \Eq{diffrate_manip1-nuc}) 
\begin{equation}\label{eq:diffrate_ep}
\frac{\ud R_N}{\ud E'} = \epsilon(E') \int_0^\infty \ud \ER \, G_N(E',\ER) \, \frac{\ud R_N}{\ud \ER}~.
\end{equation}
Thus, summing over all nuclides in the target the total rate can be written in the form 
\begin{equation}\label{diffrate_manip1}
\frac{\ud R}{\ud E'} = \frac{\sigma_\text{ref} \rho}{m} \int_{v \geqslant v_\delta} \ud^3 v \, \frac{f_\chi (\vec{v},t)}{v} \, \frac{\ud \mathcal{H}}{\ud E'} (\vec{v}, E') \, ,
\end{equation}
where the response function  $\ud \mathcal{H}/{\ud E'}= \sum_N \ud \mathcal{H}_N/{\ud E'}$ (\Eq{eq:dHcurlTotal'-gen}) is~\cite{DelNobile:2013cva,Gelmini:2015voa,Gelmini:2016pei}

\begin{equation}\label{eq:dHcurl}
 \frac{\ud \mathcal{H}}{\ud E'}(\vec{v}, E') \equiv 
  \begin{dcases} 
      \hfill \sum_N N_N\int_{\ER^{-}}^{\ER^{+}} \, \ud \ER \,\epsilon(E',\ER) \, G_N(\ER, E') \, \frac{v^2}{\sigma_\text{ref}} \, \frac{\ud \sigma_N}{\ud \ER}(\ER, \vec{v})    \hfill & \text{ if $v \geqslant v_\delta$,} \\
      \hfill 0 \hfill & \text{ if $v < v_\delta$.} 
  \end{dcases}
\end{equation}
Here, $N_N$ is the number of nuclei per unit mass of the detector,
$\sigma_N$ is the DM-nucleus scattering cross section and $v_\delta$ is the smallest of the speed threshold values $v_\delta^N$ for all nuclides in the target. Notice the formal similarity between \Eq{eq:dHcurl} and \Eq{CurlyH-App}. Similarly to the response function in \Eq{CurlyH-App},   the response function for nuclear scattering in \Eq{eq:dHcurl} depends only on the speed $v$, instead of the velocity vector $\vec{v}$ if the scattering does not depend on the direction of the incoming DM particle. 

As for DM scattering off electrons, also for inelastic endothermic scattering off nuclei for which $v_\delta \not= 0$ (these are \Eq{drate_detatilde-nuc} to \Eq{diffrate_eta} applied to this type of DM) 
\begin{equation}\label{drate_detatilde}
\frac{\ud R}{\ud E'} = - \int_{v_\delta}^{\infty} \ud v \, \frac{\partial \tilde{\eta}(v,t)}{\partial v} \, \frac{\ud \mathcal{H}}{\ud E'}(v, E') \, ,
\end{equation}
and, using that $\tilde{\eta}(\infty,t) = 0$ and $\ud \mathcal{H} / \ud E' (E', v_\delta) = 0$, integrating \Eq{drate_detatilde} by parts  results in 
\begin{equation}
\frac{\ud R}{\ud E'} = \int_{v_\delta}^{\infty} \ud \vmin \tilde{\eta}(\vmin, t) \, \frac{\ud \mathcal{R}}{\ud E'}(\vmin, E') \, ,
\end{equation}
where the differential response function $\ud \mathcal{R}/ \ud E'$ is
\begin{equation}
\frac{\ud \mathcal{R}}{\ud E'}(\vmin,E') \equiv \frac{\partial}{\partial \vmin}\left[ \frac{\ud \mathcal{H}}{\ud E'}(\vmin, E') \right] \, .
\end{equation}
As for DM scattering off electrons, here
if $v_\delta$ is in the observable range of an experiment, the response function ${\ud \mathcal{R}}{\ud E'}$ depends on a Jacobian factor 
$\partial{\ER}^\pm/\partial \ER$ which is singular at $v=v_\delta$.

\clearpage
\bibliography{halo}
\addcontentsline{toc}{section}{Bibliography}
\bibliographystyle{JHEP}

\end{document}